\documentclass{WileyMSP-template}
\usepackage[usenames,dvipsnames]{xcolor}
\usepackage[utf8]{inputenc}
\usepackage{siunitx}
\usepackage{subcaption}

\begin{document}

\title{Antibubbles enable tunable payload release with low-intensity ultrasound}

\maketitle


\author{Nicolas Moreno-Gomez*$^\dagger$}
\author{Athanasios G. Athanassiadis*$^\dagger$}
\author{Albert T. Poortinga}
\author{Peer Fischer}

$^\dagger$ These authors contributed equally.

\dedication{}

\begin{affiliations}
Dr. Nicolas Moreno-Gomez\\
Address: Institute for Molecular Systems Engineering and Advanced Materials, Im Neuenheimer Feld 225, 69120 Heidelberg, Germany and Max Planck Institute for Medical Research, Jahnstr. 29, 69120 Heidelberg, Germany.\\
Email Address: nmoreno@uni-heidelberg.de\\\hfill \break

Dr. Athanasios G. Athanassiadis\\
Address: Institute for Molecular Systems Engineering and Advanced Materials, Im Neuenheimer Feld 225, 69120 Heidelberg, Germany and Max Planck Institute for Medical Research, Jahnstr. 29, 69120 Heidelberg, Germany.\\
Email Address: thanasi@uni-heidelberg.de\\\hfill \break

Dr. Albert T. Poortinga\\
Address: Polymer Technology group, Eindhoven University of Technology, Eindhoven, The Netherlands\\
De Rondom 70, 5612 AP Eindhoven, The Netherlands\\\hfill \break

Prof. Dr. Peer Fischer\\
Address: Institute for Molecular Systems Engineering and Advanced Materials, Im Neuenheimer Feld 225, 69120 Heidelberg, Germany and Max Planck Institute for Medical Research, Jahnstr. 29, 69120 Heidelberg, Germany.\\
\end{affiliations}\hfill \break


\hfill \break

\begin{abstract}
The benefits of ultrasound are its ease-of-use and its ability to precisely deliver energy in opaque and complex media. However, most materials responsive to ultrasound show a weak response, requiring the use of high powers, which are associated with undesirable streaming, cavitation, or temperature rise. These effects hinder response control and may even cause damage to the medium where the ultrasound is applied. Moreover, materials that are currently in use rely on all-or-nothing effects, limiting the ability to fine-tune the response of the material on the fly. For these reasons, there is a need for materials that can respond to low intensity ultrasound with programmable responses. Here it is demonstrated that antibubbles are a low-intensity-ultrasound-responsive material system that can controllably release a payload using acoustic pressures in the kPa range. Varying their size and composition tunes the release pressure, and the response can be switched between a single release and stepwise release across multiple ultrasound pulses. Observations using confocal and high-speed microscopy revealed different ways that can lead to release. These findings lay the groundwork to design antibubbles that controllably respond to low-intensity ultrasound, opening a wide range of applications ranging from ultrasound-responsive material systems to carriers for targeted delivery.
\end{abstract}


\section{Introduction}

Ultrasound is a versatile tool in diverse fields, ranging from chemistry and fabrication\cite{melde2018acoustic, aubert2019spatial, zhou2022advances}, to biology and medicine \cite{blackmore2023ultrasound}. Recent advances in ultrasound-responsive materials as well as advances in physical acoustics have expanded the settings in which ultrasound can be used and the capabilities that it can provide \cite{athanassiadis2021ultrasound,li2022recent}. In particular, the ability to precisely shape ultrasound fields \cite{melde2016holograms} has made it possible to remotely trigger responses in localized regions on demand \cite{aubert2019spatial}, while benefiting from ultrasound's unique characteristics compared to other methods of remote stimulation. For instance, ultrasound provides higher spatial resolution than magnetics,  while propagating further than light through complex and opaque media \cite{lin2021magnetism,wang2017external,dholakia:2020}. Additionally, ultrasound can drive different physical phenomena -- via acoustic radiation forces, cavitation or streaming -- that can be exploited when developing new responsive systems \cite{athanassiadis2021ultrasound}. Nevertheless, despite the benefits of ultrasound, there are few materials that can be triggered or controlled with ultrasound. \hfill \break

Most ultrasound-responsive materials require moderate- or high-intensity ultrasound to trigger a desired effect \cite{athanassiadis2021ultrasound,rwei2017ultrasound}. Examples of this mostly arise in controlled drug delivery, where the carrier, often a microbubble, is stimulated using ultrasound to release a drug and initiate a therapeutic effect \cite{delaney2022making}. Beyond drug delivery there are examples where ultrasound-responsive systems are used to induce self-healing in construction materials \cite{xu2021employing,song2022influence} or for in-vivo fabrication of hydrogels through external ultrasound stimulation \cite{nele2020ultrasound}, demonstrating the importance of controlling the payload release for different applications. In many examples, high-intensity ultrasound is required to release a payload from a carrier such as microbubbles \cite{stride2019nucleation}, liposomes \cite{afadzi2012effect,el2021molecular}, or phase-change droplets \cite{couture2011ultrasound} using acoustic pressures on the order of MPa and frequencies of tens of kHz up to a few MHz \cite{athanassiadis2021ultrasound,delaney2022making, chomas:2000, chomas:2001,shapiro:2014,chandan:2020}. \hfill \break

However, the use of high-intensity ultrasound can lead to undesirable effects that limit its application in sensitive material or biological systems. Thermal, mechanical, and chemical damage is possible within the ultrasound beam through absorption, cavitation, and sonochemical effects. The onset of such effects are dependent on a combination of intensity (pressure), frequency, and exposure time. Commonly adopted thresholds for high-intensity ultrasound correspond to pressures of \SI{150}{\kilo\pascal} (time-average) and \SI{2}{\mega\pascal} (peak) at \SI{1}{\mega\hertz} in water \cite{usfoodanddrugadministration:2023}. However, detrimental effects in materials have been observed at pressures as low as \SI{170}{\kilo\pascal} at \SI{1}{\mega\hertz} \cite{thanhnguyen:2017}, and certain systems can be much more susceptible to damage than others (e.g. in ophthalmology \cite{usfoodanddrugadministration:2023, duck:2007}). In order to be useful in a wide range of contexts without inducing unwanted damage, there is therefore a need for ultrasound-responsive materials that respond to low-intensity ultrasound, ideally in the range of \SI{200}{\kilo\pascal} or below.\hfill \break

An additional challenge when using existing ultrasound-responsive systems is their release response. There is typically a narrow range of excitation pressures, above which the carrier is destroyed and the payload is released in a single event \cite{wang2017external,stride2019nucleation,baresch2020acoustic}. While such behavior might be desirable in some applications, e.g. for rapid delivery of a payload, there are other settings where more control is required. For instance, when the payload should be precisely dosed in response to real-time feedback \cite{rwei2017ultrasound}, or when the payload needs to be slowly delivered over an extended period of time \cite{kar2022wearable}. In such settings, existing ultrasound-responsive carriers do not provide adequate solutions. \hfill \break

In this work, we identify Pickering-stabilized antibubbles as a new carrier for triggered release and show that they respond robustly to low-intensity ultrasound. Antibubbles consist of one \cite{poortinga2011long,silpe2013generation,jiang2023high} or more liquid droplets \cite{poortinga2013micron,kotopoulis2022formulation} surrounded by a gas layer. This structure is not naturally thermodynamically stable \cite{vitry2019controlling}, but can be stabilized using fumed silica particles (Pickering stabilization) instead of surfactant molecules \cite{poortinga2011long,poortinga2013micron,zia2022advances}. Unlike microbubble-based carriers, which require specially-modified payloads that can be attached to the outer shell \cite{lentacker2009drug}, the internal droplets in an antibubble can carry large volumes of payload without special preparation. Because of these benefits, antibubbles are a new class of ultrasound contrast agents that have recently been proposed for ultrasonic drug delivery \cite{kotopoulis2022formulation}.\hfill \break

Here we demonstrate, for the first time, that payload delivery via antibubbles can be triggered by low-intensity ultrasound with controllable temporal release profiles. Moreover, we show that the release response can be shifted to lower pressures, and can exhibit different qualitative behaviors by tuning the formulation. We use optical fluorescence measurements to detect the release of the payload as a function of ultrasound exposure and antibubble composition. We find that, unlike with existing carriers, it is possible to change the release behavior from single release to multiple-release by varying the composition of the antibubble. Similarly, the release pressure can be adjusted via composition variation. Finally, we use high-speed brightfield microscopy to explore the origins of the different release mechanisms. Our results open the door to designing tailored antibubble formulations for specific applications, making antibubbles a valuable and versatile component for the design of new low-intensity ultrasound-activated smart materials.\hfill \break


\section{Results and Discussion}

\subsection{Antibubble Fabrication and Characterization}

\begin{figure}
\centering
  \includegraphics[scale=0.7]{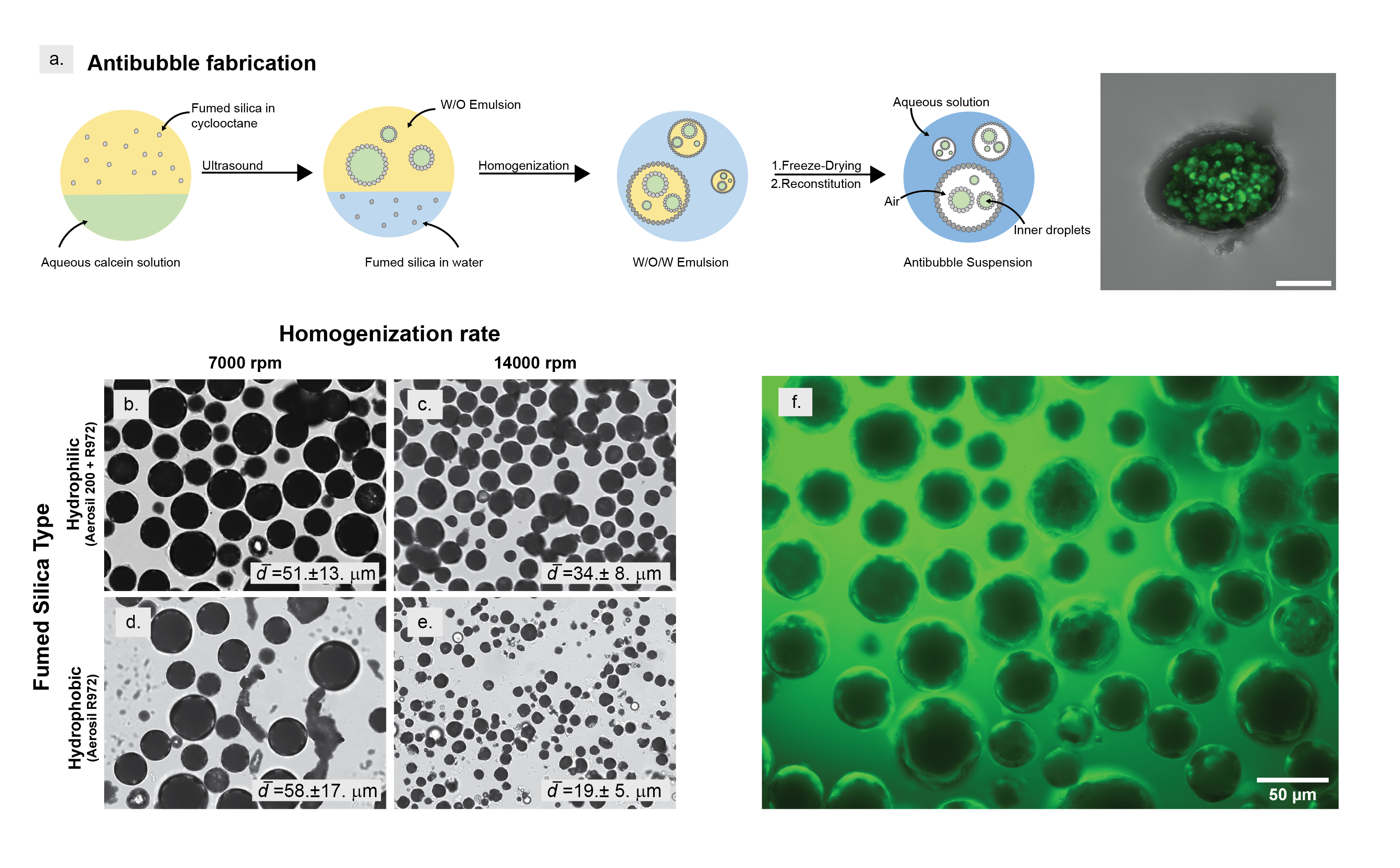}
  \caption{Fabrication and characterization of antibubbles. (a) Antibubbles were fabricated by freeze-drying a double emulsion template with multiple cores containing calcein, as visualized using Confocal microscopy (scale bar corresponds to \SI{15.}{\micro\meter}. (b)-(e) Both homogenization rate and fumed silica composition play a role in determining the size of the antibubbles. Hydrophilically-coated antibubbles obtained after homogenization (Rotor-stator) at (b) 7000 rpm  and (c) 14000 rpm show a small decrease in size with shear rate, while hydrophobically-coated antibubbles obtained after homogenization at (d) 7000 rpm and (e) 14000 rpm vary in size more significantly. (f) In fluorescence images the reconstituted antibubbles (here: hydrophilic, \SI{51}{\micro\meter}) can be identified by their dark cores surrounded by a gas layer. The dark cores reflect the quenched calcein, while free calcein in the background solution provides the ambient green fluorescence. }
  \label{fig:BF_Antibubbles}
\end{figure}

Antibubbles are fabricated \cite{kotopoulis2022formulation} by creating a water-in-oil-in-water (W/O/W) Pickering emulsion template with the desired payload dissolved in the inner water phase. The solvent is then extracted by freeze-drying, creating a stable powder template with a very long shelf life. Before use, the antibubbles are reconstituted from the powder in an aqueous solution of sodium chloride. The fabrication process is schematically illustrated in Figure \ref{fig:BF_Antibubbles}a along with a confocal image of a single antibubble to illustrate their structure (also see video S1). For this image, we incorporate a fluorescent calcein dye as a payload at concentration of 2.6 mM in order to visualize the core structure. These images confirm that multiple droplets are encapsulated in the core, and that these tend to sediment toward the bottom of the antibubble because of their higher density compared to the gas phase \cite{silpe2013generation}.\hfill \break

As has been previously described for other carriers \cite{afadzi2012effect}, to study and quantify the release from antibubbles with ultrasound, we increase the concentration of the fluorescent calcein to 119 mM  to use the self-quenching of calcein at high concentrations (Figure S1) to distinguish between intact antibubbles which are characterized by low fluorescence levels when the dye is encapsulated, and antibubbles from which a payload is released. After release, the calcein is diluted in the surrounding fluid, quenching ceases, and fluorescence is observed. \hfill \break

The calcein-loaded inner phase of the antibubble template is prepared in an aqueous solution and also includes sodium hydroxide to enhance the calcein solubility, as well as maltodextrin as a stabilizer for freeze-drying. Small droplets of concentrated calcein are then formed  by combining the aqueous solution with a dispersion of hydrophobic fumed silica in cyclooctane. This water-in-oil (W/O) emulsion is formed by vigorously stirring the two components together using an ultrasonic homogenizer (unrelated to the  ultrasonic transducers used in the experiments below). The W/O emulsion is further homogenized with an aqueous dispersion of fumed silica, maltodextrin, mannitol and sodium chloride to produce the final water-in-oil-in-water (W/O/W) template. The additives are included to balance the osmotic pressure, thereby preventing the inner droplets from swelling or deswelling. The W/O/W template is formed by rotor-stator homogenization. It is important to note that the final size of the antibubbles is largely dictated by the homogenization rate used in this step. Next, the template is washed and concentrated using centrifugation: the positively-buoyant W/O/W droplets rise to the top without coalescing, and can be separated from any free (unencapsulated) calcein. The concentrates obtained after centrifugation are freeze-dried to replace the intermediate oil phase with air. During this step, it is important to avoid structure collapse that can occur if the sample melts before starting the drying process. Therefore, we freeze the sample at temperatures below the glass transition temperature, which in our case is controlled by the concentration of maltodextrin in the internal and external aqueous phases. The concentrates were frozen in a round bottom glass flasks for 6 h in a freezer at -60$^{\circ}$C, a temperature low enough to prevent melting of the structure before starting the drying process. The samples were then dried by connecting the round-bottom flasks to a 0.1 mbar vacuum line for 24 h. During this time, the cyclooctane and most of the water are removed to produce a dried powder that can be stored for long periods of time. Finally, antibubbles are formed by reconstitution of the powder in an aqueous solution of sodium chloride.\hfill \break

We fabricated and characterized different antibubble formulations to test their response to ultrasound. In analogy to microbubbles \cite{borden2018reverse}, we hypothesized that size and interfacial energy dictate the acoustic response of antibubbles. Therefore, four different formulations were prepared by varying the size of the template and the composition of the outer shell, while keeping the inner cores the same for all formulations. The sizes are varied by homogenizing the W/O/W emulsion at either 7000 or 14000 rpm. The outer shell is varied by using two distinct types of commercially-available fumed silica particles. One type (Aerosil 200) was predominantly hydrophilic, while the other (Aerosil R972) has a modified surface with carbon content of 0.6-1.2$\%$ \cite{EvonikAerosil}, making it predominantly hydrophobic. The first formulation we use in experiments consists of the hydrophilic Aerosil 200 in a 2:1 mass ratio with Aerosil R972, and is thus referred to as `hydrophilic' below. The second formulation consists of only hydrophobic Aerosil R972 fumed silica and is referred to as `hydrophobic' below. As shown in Figures \ref{fig:BF_Antibubbles}b-e, both formulations produce a similar size distribution when homogenized at 7000 rpm. However, at 14000 rpm the hydrophobic formulation produces much smaller antibubbles. All formulations show good stability over time (see SI Figure S2) by retaining their size for a period of 20 h. Quenching of inner fluorescence is indicated by the appearance of the dark antibubble centers in both brightfield and fluorescence microscopy (Figure \ref{fig:BF_Antibubbles}f and supporting video S2). Free calcein in the external aqueous phase is removed via further washing of the antibubbles: the buoyant antibubbles are collected from the top of the aqueous solution and transferred to a new sodium chloride solution during the washing steps.\hfill \break

\subsection{Release Triggered by Low Intensity Ultrasound}

\begin{figure}
  \centering
  \includegraphics[scale=1]{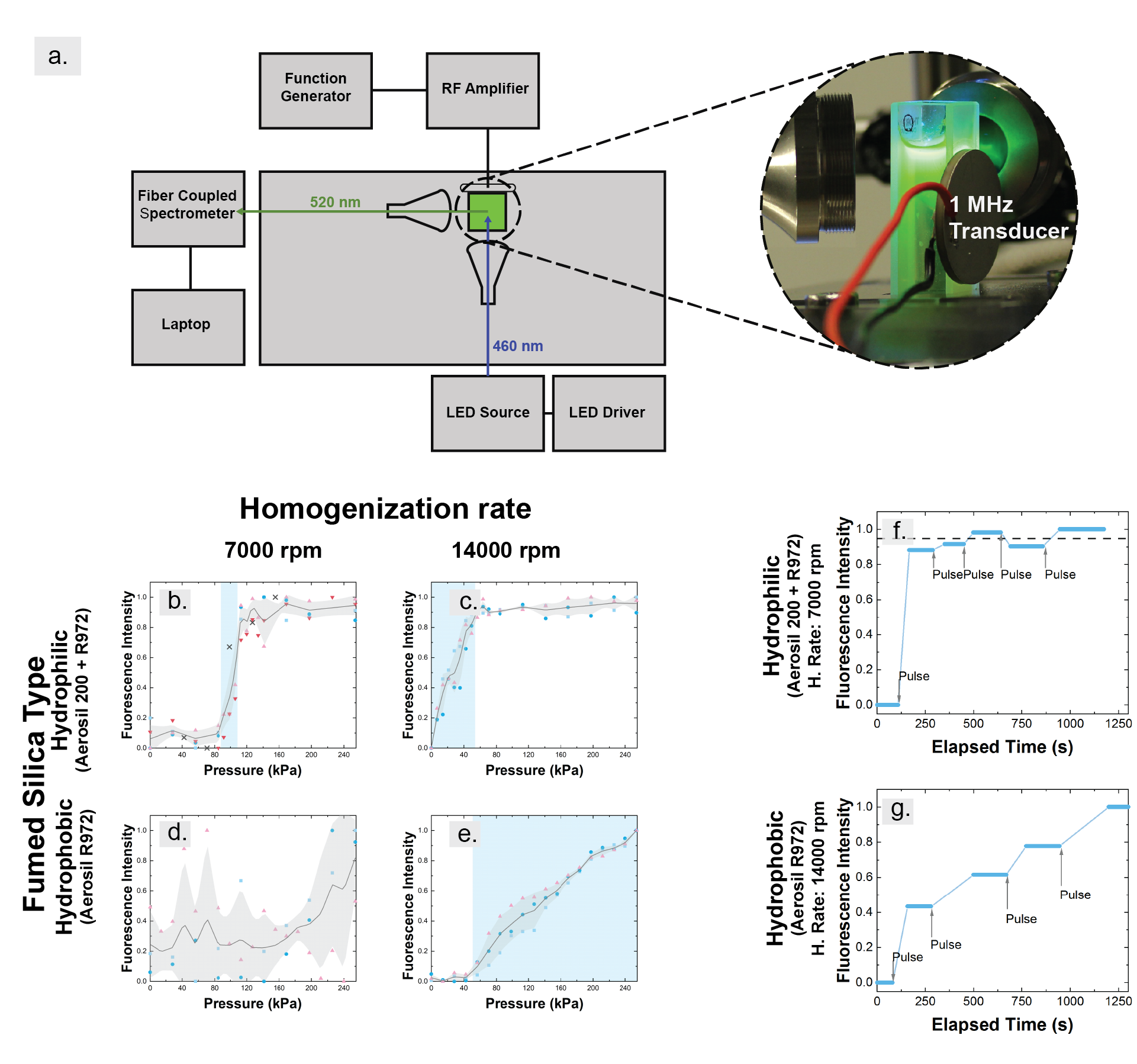}
  \caption{Experimental setup and results for fluorescence measurements as a function of ultrasound exposure. (a) Antibubbles were loaded into a cuvette and exposed to ultrasound using an integrated transducer at 1 MHz. Fluorescence signals at 520 nm were measured using a fiber-coupled spectrometer, oriented perpendicular to the excitation light (460 nm). (b)-(e) The antibubble response to ultrasonic pressure depends on antibubble composition. The pressure response for (b) \SI{51}{\micro\meter} and (c) \SI{34}{\micro\meter} hydrophilic antibubbles both demonstrate a clear threshold behavior with fluorescence intensities saturating above the threshold. The response of (d) \SI{58}{\micro\meter}  and (e) \SI{19}{\micro\meter} hydrophobic antibubbles show differing behaviors, with a low-intensity threshold only observable in the smaller antibubbles. Markers represent measurements on independent samples, the solid line represents the averaged intensity and the shading indicates one standard deviation. (f) The hydrophilic antibubbles release all the fluorescence after a single ultrasound pulse, with no significant release after subsequent pulses at the same pressure (140 kPa). (g) The hydrophobic antibubbles release the core fluorescence incrementally across multiple pulses at the same pressure (70 kPa).\hfill \break
  }
  \label{fig:setup_release}
\end{figure}

To determine the response of antibubbles to ultrasound, we measured the change in fluorescence caused by the release of calcein after exposure to ultrasound. The experimental setup is shown in Figure \ref{fig:setup_release}a. Antibubbles were observed in a  quartz spectroscopy cuvette that had an ultrasonic transducer bonded to one side of the cuvette. Fluorescence was measured by illuminating the sample with 460 nm light from a fiber-coupled LED, and collecting the emitted light with a fiber-coupled spectrometer oriented 90$^\circ$ to the excitation fiber. The fluorescence intensity at 520 nm, corresponding to the calcein emission peak, was used as an indicator of antibubble bursting.\hfill \break

Because the antibubbles are positively buoyant, the sample was mixed by gently shaking the cuvette before each fluorescence measurement and ultrasound exposure. Separate measurements confirmed that the antibubbles do not burst during gentle shaking. Before any ultrasound exposure, a baseline fluorescence level was measured. Then, the sample was gently shaken and exposed to an ultrasound pulse at 1 MHz for 2 s. This process was repeated with increasing ultrasound pressures to identify when the antibubbles release the payload. Fluorescence signals were collected continuously during and for 1 min after each ultrasound exposure to confirm the signal stability. The ultrasound response was measured using at least three different samples of each antibubble formulation. Because absolute fluorescence intensities vary from sample to sample, the resulting signal was normalized after background subtraction (see SI Figure S3 for raw fluorescence data). The normalized fluorescence intensity is plotted as a function of acoustic pressure for each formulation in Figures \ref{fig:setup_release}b-e. Our measurements reveal that the antibubble response to ultrasound is strongly dependent on the outer shell composition and size. The results are plotted in Figure \ref{fig:setup_release} and summarized in Table \ref{tab:ABResponseSummary}. This opens up the possibility to tune the release characteristics in these carriers. \hfill \break

Antibubbles with hydrophilic outer shells demonstrate single-release behavior, with the release pressure dependent on the size. The larger (\SI{51}{\micro\meter}) hydrophilic antibubbles (Figure \ref{fig:setup_release}b) release calcein already at a low pressure of \SI{88}{\kilo\pascal} (equivalent intensity \SI{258}{\milli\watt\per\centi\meter\squared}), with fluorescence release increasing to saturation at \SI{109}{\kilo\pascal}. The smaller (\SI{34.}{\micro\meter}) antibubbles (Figure \ref{fig:setup_release}c) demonstrated a similar release behavior, yet starting from a lower pressure below \SI{7}{\kilo\pascal} (equivalent intensity \SI{2}{\milli\watt\per\centi\meter\squared}) , with the amount of calcein release increasing until \SI{53}{\kilo\pascal}, where the fluorescence signal saturates, indicating no further release from the core. These differences in the response indicate that by changing the size, the required acoustic pressure for release can be shifted. \hfill \break

To confirm that the observed fluorescence release was due solely to the pressure level and not to the repeated exposures of ultrasound, we repeated our measurements for select pressures with new samples for each pressure (indicated by cross markers in Figure \ref{fig:setup_release}b; 5 independent samples). The amount of fluorescence measured in these independent experiments matches closely with the previously measured values. As a final demonstration that multiple exposures do not significantly affect the measurements for hydrophilic antibubbles, we exposed a single sample of the large hydrophilic antibubbles to multiple pulses at \SI{140}{\kilo\pascal}, which is above their bursting pressure. As shown in Figure \ref{fig:setup_release}f, the fluorescence increases drastically after the first ultrasound pulse, with minimal further release of calcein after subsequent pulses. These measurements confirm that pressure plays the dominant role in determining release for the hydrophilic antibubbles, and not multiple exposures. These results also indicate that the antibubbles are stable against ultrasound exposure at pressures below the bursting pressure. \hfill \break

In contrast to the hydrophilic antibubbles, those with the hydrophobic shell demonstrate more varied size-dependent behavior in response to ultrasound. As shown in Figure \ref{fig:setup_release}d, the larger (\SI{58}{\micro\meter}) antibubbles do not release appreciable amounts of calcein after irradiation. The same conclusion can be drawn from the raw fluorescence signal (Figure S3c). The formulation with a smaller diameter (\SI{19.}{\micro\meter}), however, shows that above \SI{49}{\kilo\pascal} (equivalent intensity \SI{80}{\milli\watt\per\centi\meter\squared}) the fluorescence intensity increases linearly with each ultrasound pulse to at least \SI{250}{\kilo\pascal} (Figure \ref{fig:setup_release}e), indicating an ongoing release process that does not saturate. To test this behavior further, we measured the calcein released by these small, hydrophobic antibubbles across multiple pulses at \SI{70}{\kilo\pascal} (Figure \ref{fig:setup_release}g). We find that even at a single excitation pressure, the antibubbles release small amounts of calcein after each pulse, reflecting a stepwise release process rather than a single one. \hfill \break

\begin{table}
\caption{Response of different antibubble formulations to ultrasonic excitation at 1 MHz.
\label{tab:ABResponseSummary}}
\centering
 \begin{tabular}[htb]{@{}ccccc@{}}
    \hline
    Shell Composition$^\dagger$ & Homogenization Rate & Size & Release Type & Release Threshold \\
    \hline
    Hydrophilic & \SI{14000}{rpm}  & \SI{34(8)}{\micro\meter} & Single & \SI{<7}{\kilo\pascal} \\
    Hydrophilic & \SI{7000}{rpm}  & \SI{51(13)}{\micro\meter} & Single & \SI{88(7)}{\kilo\pascal} \\
    Hydrophobic & \SI{14000}{rpm}  & \SI{19(5)}{\micro\meter} & Multi & \SI{49(14)}{\kilo\pascal} \\
    Hydrophobic & \SI{7000}{rpm}  & \SI{58(17)}{\micro\meter} & None & \SI{> 250}{\kilo\pascal} \\
  \hline
  \multicolumn{5}{l}{\footnotesize $\dagger$ Hydrophobic - Aerosil R927; Hydrophilic - Aerosil 200 and Aerosil R972 (2:1)} \\
  
\end{tabular}
\end{table}

In addition to the fluorescence measurements, we checked how ultrasound exposure alters the size distribution of antibubbles. We sampled hydrophilic antibubbles from a single batch before and after exposure to \SI{140}{\kilo\pascal} ultrasound and measured their sizes using brightfield microscopy and standard image processing tools (SI Figure S4). These measurements show that not all antibubbles burst. Rather, the mean diameter of the antibubbles shifts from \SI{51}{\micro\meter} to \SI{45}{\micro\meter} ($P<0.05$, $n=300$). In addition, the fraction of antibubbles larger than \SI{50}{\micro\meter} decreased while the fraction of smaller antibubbles increased. This observation suggests that antibubbles with a diameter larger than \SI{50}{\micro\meter} are more sensitive to the \SI{1}{\mega\hertz} ultrasound. However, given the observations of multiple release in the smaller hydrophilic antibubbles it is still unclear whether the smaller hydrophobic antibubbles can release calcein without being completely destroyed. Similarly, it is not clear whether the change in size distribution is due to complete destruction of larger antibubbles, or fragmentation that produces smaller bubbles. \hfill \break

Our results clearly demonstrate that the response of antibubbles to ultrasound is strongly dependent on outer shell composition and antibubble size. While the hydrophilic antibubbles release the core contents in a single bursting event upon ultrasound exposure, the hydrophobic antibubbles release less of their core contents per exposure, providing on-demand dosing of the payload. In both cases, the bubble sizes determine the pressure threshold, above which the antibubbles begin releasing their contents. This behavior, however also shows different trends depending on the shell composition. For the hydrophilic antibubbles, smaller bubbles exhibit a higher release threshold, while for the hydrophobic antibubbles, the larger bubbles do not exhibit any release behavior within the range of pressures investigated. These opposing trends raise interesting questions for the role of size in antibubble release. By analogy with microbubbles, one might expect resonant effects to play an important role, with bubbles resonant closer to the excitation frequency requiring lower pressures for release than bubbles further from resonance. However by this logic the smaller antibubbles should have released at lower pressures compared with the larger ones when excited at \SI{1}{\mega\hertz} \cite{kotopoulis:2015}, which is not the case for the hydrophilic antibubbles. These results indicate a more complex interplay between size and shell composition in determining the acoustic response of antibubbles.\hfill \break

The multi-release behavior and low pressure thresholds make antibubbles more versatile tools than existing carriers. Most existing carriers release their payload all at once \cite{athanassiadis2021ultrasound}, with only a few capable of gradual release \cite{rwei2017ultrasound,noble2014digital,xia2016ultrasound,kennedy2016sequential,moncion2018sequential, aliabouzar2021stable}. One of the few examples where two different release mechanisms have been observed is in polymeric micelles. Depending on the type of polymer used and the ultrasound parameters (frequency, amplitude), single release was possible after irreversible disruption of the micelle structure, or partial release could be caused by temporal disruption of the self-assembled structure \cite{xia2016ultrasound}. Since each of these responses is caused by different phenomena, it is necessary to use different ultrasound amplitudes for each polymer structure. By contrast, antibubbles exhibit the two different behaviors under similar ultrasound conditions and only require different composition of the external shell. This makes antibubbles one of the few carriers where the release can be tuned. Interestingly, the release profile can be adjusted using the same ultrasound settings. Furthermore, much lower acoustic pressures are required to release the contents of the antibubbles compared to other carriers like liposomes (1.5-2.0 MPa)(\cite{el2021molecular}), microbubbles (0.5-5.0 MPa) \cite{stride2019nucleation} or PFC-droplets (0.3-8.5 MPa) \cite{couture2011ultrasound}. The release pressures for antibubbles (\SIlist{7;49;88}{\kPa}) correspond to low intensities (\SIlist{2;80;258}{\milli\watt\per\centi\meter\squared}) and low mechanical indices (0.01, 0.05, and 0.09 for \SI{1}{\mega\hertz} excitation). These low pressures preclude nonlinear propagation in the background medium and ensure that destructive thermal or cavitation effects do not pose a risk to surrounding media. The low release threshold is also specific to ultrasound, and the bubbles remain stable even in the presence of mild mechanical agitation such as flows and stirring. While higher background ultrasound intensities could spuriously trigger unspecific release, targeted ultrasonic fields typically localize the ultrasonic pressure sufficiently and therefore should not trigger release beyond the higher-intensity focus. Between the specificity of the the release to ultrasonic stimulation and the low release thresholds, antibubbles can be a very attractive candidate for applications requiring tunable, on-demand payload release in sensitive environments.\hfill \break

\subsection{Antibubble bursting dynamics and release mechanisms}

The ability to tune antibubble properties for use in different settings requires a better understanding of the release mechanism for individual antibubbles, and how this mechanism varies with size and composition. As discussed above, resonant effects are expected to play a role, however this is likely in a complex interplay with other effects stemming from the surface energy of the antibubbles. Other effects, such as direct stretching of the fluid and antibubbles in the acoustic waves, can be ruled out as the $\sim\SI{100}{\kilo\pascal}$ pressures used above produce a peak molecular displacement of \SI{11}{\nano\meter}. Taken over the acoustic wavelength (\SI{1.5}{\milli\meter} at \SI{1}{\mega\hertz}), this corresponds to a peak strain on the order of \SI{0.0007}{\percent} - far too small to have a direct effect on antibubble release.\hfill \break

In order to clarify what happens at a single-bubble level during ultrasound exposure, we made use two different microscopy techniques. First, we measured the fluorescence released from individual antibubbles using confocal microscopy. Second, to observe the dynamics of antibubble release, we recorded the response of antibubbles to ultrasound using high-speed brightfield microscopy. These two imaging techniques provide us complementary information on the single-antibubble release dynamics: confocal imaging allows us to assess release and bursting after an entire pulse, while the high-speed imaging reveals antibubble dynamics on short timescales associated with each ultrasound oscillation. In these experiments we only investigated the large hydrophilic antibubbles, since the better-defined pressure threshold allows for clearer insights into the release process.  \hfill \break

\begin{figure}
\centering
  \includegraphics[scale=0.5]{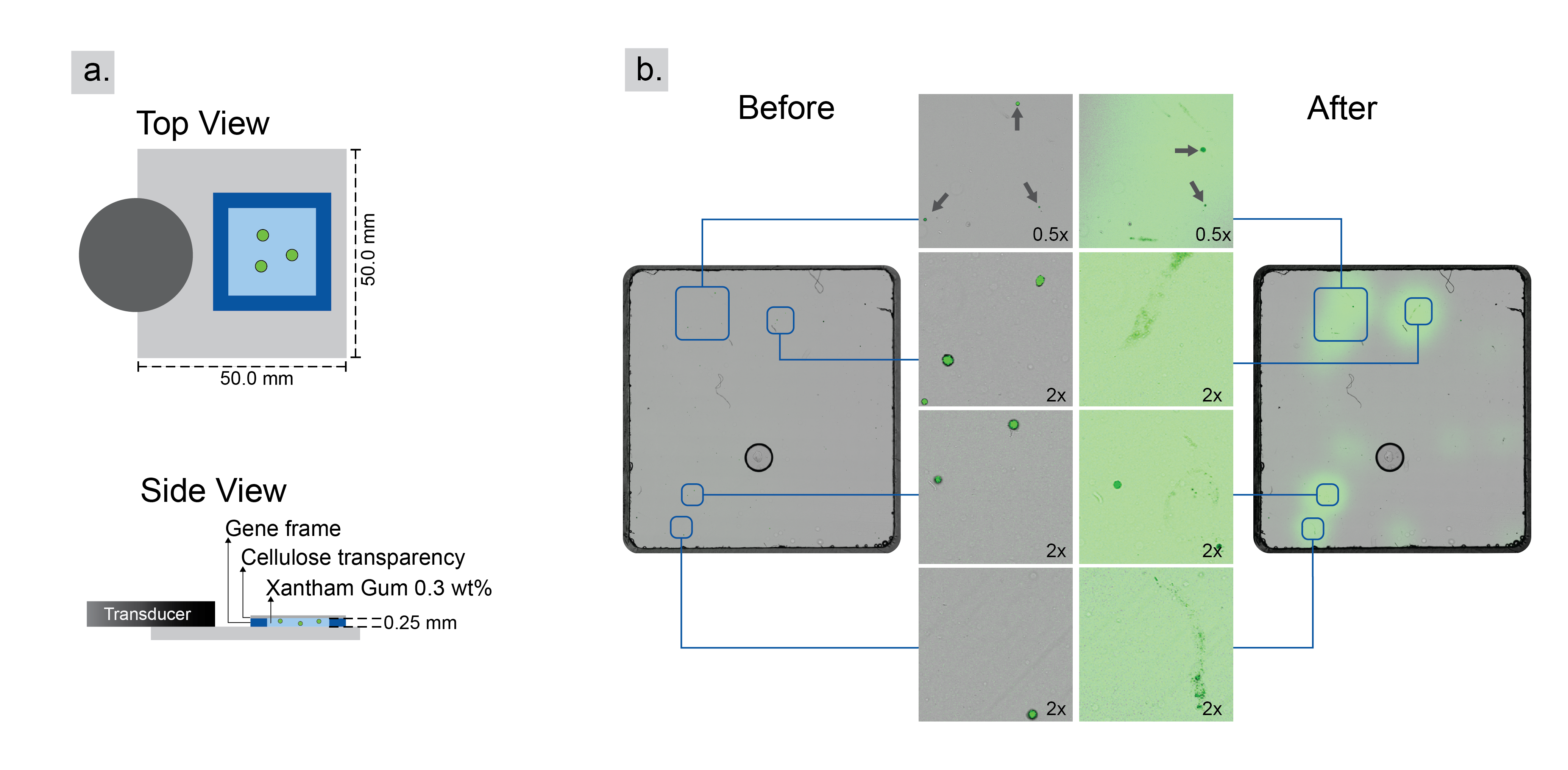}
  \caption{Observation of payload release from individual {hydrophilic} antibubbles using confocal fluorescence microscopy. Panel (a) illustrates the glass-slide setup used for the observation, where antibubbles suspended in a medium of xanthan gum are placed inside a chamber. Panel (b) shows the antibubbles before and after ultrasound stimulation. The release of the payload is confirmed by the detection of increased fluorescence locally where the antibubbles were located. {Signs of antibubble fragmentation and destruction are visible as trails of silica, alongside bubbles that have released fluorescence but otherwise remained intact.}}
  \label{fig:Confocal}
\end{figure}

For the microscopy measurements, we used a glass-plate excitation setup, which allows for transmission imaging of the sample in a small chamber (Figure \ref{fig:Confocal}a). The antibubble sample was suspended in a solution of 3. wt$\%$ of NaCl 0.3 wt$\%$ xantham gum  to minimize antibubble motion and aggregation during the excitation. A low-concentration sample was loaded into a well (Gene Frame, Thermo Scientific) on a glass plate, and the well was sealed with a transparency film. The sample was excited by a piezoelectric transducer glued to the glass plate, which was driven by \SI{300}{\milli\second} pulses at the radial resonance frequency of the transducer (\SI{90.5}{\kilo\hertz}). This excitation mode transmitted acoustic pulses through the glass into the sample chamber (SI Figure S5). The lower frequency was used to permit for direct observation of the dynamics via high-speed imaging. \hfill\break

The confocal measurements revealed clear release from individual antibubbles, as well as signs of fragmentation and partial release. As shown in Figure \ref{fig:Confocal}b, intact antibubbles before ultrasound exposure can be identified by their compact fluorescent cores. After a single 300 ms ultrasound pulse, calcein was visibly released and slowly diffused through the surrounding medium. A closer look at individual antibubbles shows evidence of motion in the ultrasound field and large fragments (most likely silica remaining after antibubble fragmentation) still emitting fluorescence, as well as some intact antibubbles with a weakly fluorescent core. This suggests that the antibubbles respond differently to ultrasound. While some antibubbles were completely destroyed during the pulse, others moved under the influence of radiation forces, leaving a trail of the fumed silica and releasing at least part of their payload. Thus intact bubbles or antibubbles after exposure also contribute to the observed fluorescence change, suggesting that the payload release does not require complete destruction of the hydrophilic antibubbles, but can also involve more complicated dynamics such as fragmentation or transient bridging of the core with the external fluid.  \hfill \break

\begin{figure}
  \centering
  \includegraphics[width=1\linewidth]{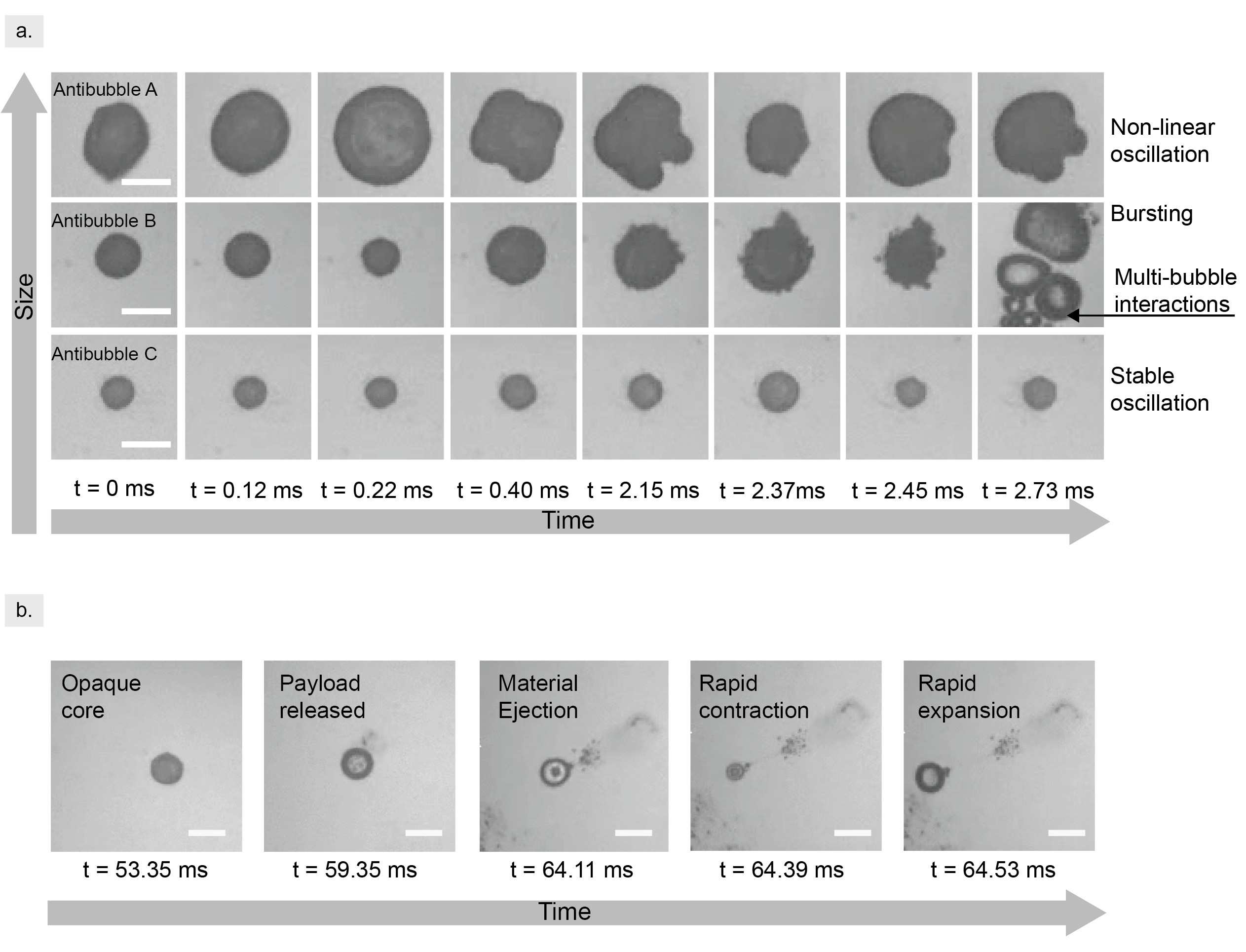}
  \caption{Time-resolved response of four antibubbles to ultrasound recorded with high-speed microscopy. In panel (a), three antibubbles of varying sizes exhibit distinct behaviors: antibubble A (\SI{72}{\micro\meter}) undergoes highly non-linear oscillations, antibubble B (\SI{47}{\micro\meter}) sheds material and interacts with an oscillating bubble cluster, and antibubble C (\SI{34}{\micro\meter}) ejects material during stable oscillations. Panel (b) shows a fourth antibubble (\SI{32}{\micro\meter}) that is similar in size to antibubble C, which releases material from the core after 15 ms, as indicated by the sudden brightening of the core due to an increase in the visibility of transmitted light through the bubble. The antibubble subsequently undergoes much larger oscillations. Frames in panel (a) were extracted from supporting Video S3 and panel (b) from Video S4. Scale bars represent \SI{50}{\micro\meter}.
  }
  \label{fig:HighSpeed}
\end{figure}

We directly observed multiple individual antibubbles using high-speed microscopy and identified three characteristic responses that could lead to payload release. The antibubble sample was again transferred to the glass plate, placed in an upright microscope and recorded with 10$\times$ magnification at 120000 frames per second using a high-speed camera. The characteristic responses are exemplified by three different antibubbles with different sizes that were observed simultaneously during ultrasound exposure (diameters: A - \SI{72}{\micro\meter}, B - \SI{47}{\micro\meter}, and C-\SI{34}{\micro\meter}). Figure \ref{fig:HighSpeed}a shows select frames extracted from supporting video S3, illustrating the response of three antibubbles. Antibubble A is the first to start oscillating, followed by antibubbles B  and C. The oscillations of antibubble A become larger with time and quickly become nonlinear, exhibiting higher order oscillations. Antibubble B, of intermediate size, oscillates until it begins to release material, probably silica, followed by its destruction after interacting with a bubble cluster, likely formed by other antibubbles out of the frame. Here, the release of the payload is not only due to the collapse of the structure, but also involves interactions with other antibubbles and with material coming from the fragmentation of other structures. Lastly, antibubble C oscillates weakly at the start of the ultrasound pulse. However, after more time (supporting video S3), the oscillations of antibubble C become larger and it releases its payload while remaining intact, revealing a third release mechanism. The same mechanism was observed in a fourth antibubble (\SI{32}{\micro\meter} diam.), shown in Figure \ref{fig:HighSpeed}b and supporting video S4. For these antibubbles, linear ultrasound-induced oscillations resulted in an eventual  release of the payload, marked by a transition from opaque dark cores to transparent ones. This process, accompanied by a slow movement of the antibubble, continues with the ejection of material (likely silica), and is followed by large oscillations that would not be possible with an intact core. Finally, the residual microbubble jets away, leaving a trace similar to what was observed in the confocal images (Figure \ref{fig:Confocal}b). \hfill \break

Our microscopy results reveal that individual antibubbles can demonstrate a rich variety of behaviors leading to payload release. Antibubbles can release during a destructive collapse or fragmentation, which are processes likely driven by large nonlinear oscillations or interactions with other bubbles or antibubbles. The fact that nonlinear oscillations can onset so quickly in low pressures is likely a result of two factors. Most importantly is the role of the core. Since the core is incompressible and typically around 80\% of the antibubble diameter, there is a smaller volume of air that can participate in oscillations, and interactions between the inner and outer interfaces of the antibubble can take place faster. Such behavior would lead to nonlinearities quickly, similar to how coatings on microbubbles have led to nonlinear `compression-only' responses \cite{sijl:2011}. Another effect that likely plays a role is resonance. Like microbubbles, antibubbles exhibit acoustic resonances \cite{kotopoulis:2015}, and an antibubble driven near its acoustic resonance will oscillate more strongly. In our microscopy experiments, the driving frequency of \SI{90.5}{\kilo\hertz} corresponds to a resonant antibubble diameter on the order of \SI{100}{\micro\meter} \cite{kotopoulis:2015}. The specific resonant radius also depends on the core diameter, which can vary slightly within a sample of antibubbles. However, this radius is clearly larger than the antibubbles in our measurements, so one would expect that the larger antibubbles would respond more strongly, and more nonlinearly, to the ultrasound in our experiments, just as we observed. \hfill\break

Beyond the destructive processes of collapse or fragmentation, our measurements also reveal a more gentle mechanism of release for the smaller antibubbles. These bubbles do not significantly change shape but clearly empty their core contents into the surrounding fluid, as indicated by the change in transparency of the antibubble. The following dynamics are also characteristic of microbubbles without a core, further supporting that they have released the payload. The specific mechanism for gentle release remains unclear, but our hypothesis is that small volumes of payload are ejected from the core by a transient bridging of the inner and outer fluids, reminiscent of `kiss and run' exocytosis \cite{fesce:1994} in cellular systems. Such effects could be enhanced by the fact that multiple droplets are encapsulated in the air bubble, so that individual core droplets could be released without affecting others. Such behavior could also play a role in the stepwise release observed in the hydrophobic antibubbles, where some core contents are retained between pulses. While our current measurement setup does not allow for high-speed fluorescence measurements to confirm these behaviors, we speculate that the hydrophobic shell stabilizes the structure in the presence of ultrasonic disruption, and would quickly close any bridges between the inner and outer fluids, so that only a small amount of the payload, or a single droplet from the multi-droplet core,  can be released at a given time. This mechanism, however, requires further investigation. \hfill\break

\section{Conclusion}

We have demonstrated that antibubbles respond to low-intensity ultrasound by releasing their payload in a controllable manner. By varying the antibubble size and composition, the release pressure can be tuned within the 1-100 \si{\kilo\pascal} range, and different release profiles can be achieved. A nontrivial interplay of the antibubble composition and size leads to different behaviors. For antibubbles with hydrophilic shells, the release pressure is larger for smaller antibubbles, while the opposite is true for hydrophobic antibubbles. Hydrophilic antibubbles displayed a characteristically sharp pressure release threshold, and ejected the majority of the payload after a single ultrasound pulse. By contrast, the smaller hydrophobic antibubbles exhibited a unique stepwise release profile, making it possible to precisely dose the payload delivery over time. A surprising result of our experiments is that the antibubbles demonstrate robust low-intensity release behavior over a wide range of frequencies ($<$\SI{100}{\kilo\hertz}) to (\SI{1}{\mega\hertz}). Unlike conventional bubble-based release, which requires intense, resonant excitation to rupture the bubble, the responsiveness of antibubbles across a broad frequency range could be advantageous in a wider range of applications. The low release threshold is also specific to ultrasound, and the bubbles remain stable even in the presence of mild mechanical agitation such as flows and stirring. While higher background ultrasound intensities could spuriously trigger unspecific release, targeted ultrasonic fields typically localize the ultrasonic pressure sufficiently and therefore should not trigger release beyond the higher-intensity focus.
We have identified multiple distinct mechanisms that could play a role in the release process under different conditions: nonlinear oscillations, multi-bubble interactions, and transient bridging. {Microscopy data shows that these mechanisms can lead to a total release of the core contents from hydrophilic antibubbles in some cases, while possibly leaving partially-filled or empty microbubbles behind in others. It remains an open question to identify if there are conditions that lead to total destruction of an entire sample and complete gas dissolution.}
 Further work is needed to understand the specific role of each mechanism at different frequencies and for specific antibubble compositions. \hfill\break

The ability to deliver payloads at low pressures, as well as the ability to tune the delivery profile between single and stepwise release, make antibubbles a unique carrier for payload delivery in diverse settings. Our results lay the foundation for designing tailor-made formulations for specific applications such as controlled drug delivery \cite{freiberg:2004, timko:2014}, acoustic fabrication \cite{melde2018acoustic}, and healing materials \cite{moncion2018sequential}. More complex payloads including hydrophobic drugs, enzymes and even simple organisms \cite{mardanighahfarokhi:2020} can be encapsulated without damage, supporting a rich set of delivery applications. In many of these cases antibubble sizes similar to those studied in this work are directly relevant, while others - primarily clinical - will require smaller antibubbles (e.g. below \SI{10}{\micro\meter}). This motivates further study of different shell compositions and investigations of different encapsulation techniques to provide more control over sample sizes and monodispersity. Further work in these directions could make it possible to build a library of antibubbles whose response behavior can be selectively addressed by pressure and frequency. Such a library would make it possible to combine distinct antibubble formulations carrying different payloads, and allow for specific triggering of single classes of antibubbles at a time, opening the door to controlling complex biological or chemical processes using antibubble carriers and low-intensity-ultrasound. \hfill\break

\section{Experimental Section}
\subsection{Antibubble fabrication}

The general fabrication of the different antibubbles was performed as described elsewhere \cite{kotopoulis2022formulation}. First, the inner aqueous solution was prepared by dissolving in 3.44 g type I water (Millipore), 0.6 g NaOH 2 M, 0.3 g calcein and 0.44 g maltodextrin. The final calcein concentration is around 119 mM. Second, 0.8 g of fumed silica Aerosil R972 (Evonik) was dispersed in 16.4 g of cyclooctane using an ultrasonic sonotrode (Hirscher, UP100) until no visible fumed silica was observed, forming the intermediate oil phase. For the external aqueous phase, 160 g of water, 20 g of maltodextrin, 20 g of mannitol and 2 g of NaCl were combined. Then 100 g of this solution was used to disperse either Aerosil R972 fumed silica alone or a mixture of Aerosil 200 with Aerosil R972 in a 2:1 mass ratio using rotor-stator homogenization (Ultraturrax T18, IKA) and an ultrasonic sonotrode (Hirscher, UP100). The final total fumed silica content of the outer phase is about 0.5 wt$\%$. Next, 4 g of the inner aqueous solution was dispersed in 17 g of the medium solution using an ultrasonic sonotrode (Hielscher, UP100H, amplitude 100$\%$, cycle 1, 60 s) to form a W/O emulsion. Next, 5 g of the W/O emulsion was combined with 40 g of an external solution containing 0.5 wt$\%$ of fumed silica by rotor-stator homogenization (Ultraturrax T18, IKA) to obtain W/O/W emulsions. The mixture was first homogenized at a speed of 7000 rpm for 2 min and then 20 g of sample was scooped out and transferred to a centrifuge tube. The remaining sample was further homogenized at 14000 rpm and then transferred to another centrifuge tube. Approximately 30 mL of the external solution without fumed silica was added to each batch and then centrifuged at 500 RCF for 2 minutes. The concentrate from the top was then transferred to a round bottom flask and frozen at -60$^{\circ}$C for at least 8 h and then dried at -85$^{\circ}$C and 0.1 mBar for 24 h (Martin Christ, Alpha 2-4 LSCplus). The resulting dry powder was stored and used after reconstitution in aqueous NaCl 2$\%$. \hfill \break

Aerosil R972 and Aerosil 200 were generously provided by Evonik. Maltodextrin 2DE and 12DE were provided by Roquette. Sodium hydroxide, calcein, cyclooctane, d-mannitol, and sodium chloride were purchased from Sigma-Aldrich. All reagents were used without further purification. 

\subsection{Antibubble bursting}

\subsubsection{Fluorescence-pressure plots}
For bulk fluorescence experiments, a lead zirconate titanate (PZT) disk transducer with 25 mm diameter and 2 mm thickness (PI Ceramic) was glued to the side wall of the cuvette using cyanoacrylate superglue (Loctite 401). Electrical leads were soldered to opposite sides of the transducer and the device was driven by a function generator (AFG2062, Tektronix) connected to a power amplifier (ENI 2200L). A sample of antibubbles (10 mg powder reconstituted in 4. mL of 3\% NaCl solution) was filled into a 4 mL quartz spectroscopy cuvette with an ultrasonic transducer bonded to one side. The sample was illuminated using a fiber-coupled LED (Thorlabs M470F4), and fluorescence was measured at 90 degrees using a fiber-coupled spectrometer (Ocean Optics Maya 2000).\hfill \break

Acoustic pressures were measured using a needle hydrophone (Onda HNR-0500) placed in the sound field. The voltage output from the hydrophone was converted to a pressure value using calibration curves provided by the manufacturer. Since standing waves are formed both in the cuvette and on the glass plate, we scanned the pressure fields spatially to find and report the peak pressures within an experiment.\hfill \break

The release thresholds are calculated for each shell composition based on at least $n=3$ independent experiments where fluorescence was measured after ultrasound pulses of increasing pressure. Between pulses, the sample was gently shaken to redistribute the positively-buoyant antibubbles throughout the cuvette. The release threshold is defined when the increase in normalized fluorescence for a given pressure excitation is statistically significant compared to the scatter in the low-pressure fluorescence data. We deemed an increase significant when the fluorescence measurement was at least 2 standard deviations above the average fluorescence for all lower pressures. The threshold was then taken as the mean of the pressure where the fluorescence increase was detected and the previous measurement pressure. The error bars in the threshold are set by the spacing between pressure measurement points. As shown in Fig. 2C and D, a response threshold could not always be determined. In the case of Fig. 2C, a significant response was already observed at the lowest excitation pressure tested. In the case of Fig. 2D, no response was observed to the highest pressure measured. 

\subsubsection{Glass plate experiments}

The glass plate excitation setup was designed to allow for ultrasound excitation of the antibubbles while permitting transmission optical measurements in a microscope. A PZT piezoelectric disk was glued to one face of a $50 \times 50$ mm glass slide and electrical leads were soldered to opposing sides of the piezo. A sample chamber was created on the glass slide using an adhesive Gene Frame (Thermo Fisher) with \SI{250}{\micro\meter} height. Antibubble samples were suspended in a solution 3 wt.\% NaCl and 0.3 wt\% xantham gum and loaded into the chamber. The chamber was sealed from above using a polyester transparency film. The piezo was connected to a signal generator (Tektronix AFG1062) via a 53 dB power amplifier (ENI 2200L), and driven with a single, \SI{300}{\micro\second} pulse of ultrasound at \SI{90.5}{\kilo\hertz}. At this frequency, the piezo excited plate waves in the glass slide, which formed a standing wave pattern in the sample chamber as shown in Figure S2. The peak pressure in the sample chamber was \SI{58}{\kilo\pascal} ($MI=0.19$, peak intensity \SI{0.1}{\watt\per\centi\meter\squared}).

\subsubsection{Microscopy characterization}

Antibubble size distributions were measured using an inverted optical microscope (Zeiss Observer D1) with a $5\times$ objective. Size distributions are based on microscopy images, where the software ImageJ was used to measure the diameter of antibubbles. The scale was determined using a calibration slide. \hfill \break

Confocal microscopy was performed on a Zeiss LSM 900 (inverted) laser-scanning confocal microscope. Brightfield and green fluorescence images of the entire sample chamber were acquired as multiple tiles with a pixel size of \SI{12.5}{\micro\meter}. The tiles were stitched together in postprocessing after background correction. Images were acquired before and after excitation with ultrasound. The excitation took place outside of the confocal microscope and the sample was returned to the microscope for further imaging.\hfill \break

High speed videos of antibubble responses to ultrasound were acquired at 120,000 frames per second using a Photron Nova S16 camera mounted onto an Olympus BX61 upright microscope. Antibubble samples in the glass slide excitation geometry were imaged using a $10\times$ objective, with a pixel size of \SI{2}{\micro\meter\per px}. The high speed camera recording and ultrasound exposure were triggered simultaneously via a TTL pulse. The camera recorded 50 frames before the trigger and 49950 frames after, covering the entire pulse. The videos were saved as an uncompressed .avi file and quantitative measurements were made using the open source software FIJI \cite{schindelin2012fiji}.

\subsection{Statistical Analysis}

For the comparison of size distributions before and after ultrasound exposure (Fig. S5) two samples of $n = 300$ were compared using a  Welch's t-test, using Origin software. Results were statistically different with a significance level of $\alpha=0.05$ and p-value $P = 2.539\times10^{-5}$.





\medskip
\textbf{Supporting Information} \par 
Supporting Information is available from the Wiley Online Library or from the author.

\medskip
\textbf{Acknowledgements} \par 
This work was supported by the European Research Council under the ERC Advanced Grant Agreement HOLOMAN (No. 788296). The authors gratefully acknowledge the data storage service SDS@hd supported by the Ministry of Science, Research and the Arts Baden-Württemberg (MWK) and the German Research Foundation (DFG) through grant INST 35/1314-1 FUGG and INST 35/1503-1 FUGG.

\clearpage
\renewcommand{\thefigure}{S\arabic{figure}}
\renewcommand{\thesection}{S\arabic{section}}
\renewcommand{\thetable}{S\arabic{table}}
\setcounter{figure}{0}
\setcounter{section}{0}
\setcounter{table}{0}
\title{Supporting Information for: Antibubbles enable tunable payload release with low-intensity ultrasound}

\maketitle


\author{Nicolas Moreno-Gomez*}
\author{Athanasios G. Athanassiadis*}
\author{Albert T. Poortinga}
\author{Peer Fischer}

\section{Intensity and Mechanical Index Calculations}

To facilitate more direct comparison with clinically-relevant ultrasound metrics, here we report the mechanical indices and equivalent plane wave intensities corresponding to the threshold pressure we identify in the main text. The mechanical index is calculated as \cite{athanassiadis2021ultrasound} $$MI = \frac{P}{\sqrt{f}},$$ where $P$ is the peak negative pressure in \si{\mega\pascal} and $f$ is the ultrasound frequency in \si{\mega\hertz}. \hfill\break

While intensities are often used in a clinical context, the release phenomena described here are driven by a pressure mechanism. The intensity in our standing-wave-based excitation geometries is therefore zero. For more meaningful comparison to clinical settings, we therefore calculate the intensity of a plane wave with the same peak pressure as we applied in our experiments. The intensity is therefore calculated as \cite{athanassiadis2021ultrasound} $$I=\frac{P^2}{2\rho c},$$ where $P$ is the peak pressure, $\rho$ is the fluid density, and $c$ is the fluid sound speed. For the calculations, we assume the fluid medium is water with a density of $\rho=\SI{1000}{\kilo\gram\per\meter^3}$ sound speed of $c=\SI{1500}{\meter\per\second}$.

\begin{table}[h]
    \centering
    \begin{tabular}{|c|c|c|c|}
    \hline
         Pressure & Frequency & Mechanical Index & Equivalent Plane Wave Intensity\\
         (\si{\kilo\pascal}) & (\si{\kilo\hertz}) & (\si{\mega\pascal\per\sqrt{\mega\hertz}}) & (\si{\watt\per\centi\meter\squared}) \\
         \hline
         7 & 1000 & 0.007 & 0.002 \\
         49 & 1000 & 0.049 & 0.080 \\
         88 & 1000 & 0.088 & 0.258 \\
         250 & 1000 & 0.250 & 2.083 \\
         \hline
         \hline
         58 & 90.5 & 0.19 & 0.11 \\
         \hline
    \end{tabular}
    \caption{Mechanical Index and Equivalent Plane Wave Intensity corresponding to the release thresholds and excitation pressures identified in the main text.}
    \label{tab:SI_intensity_MI}
\end{table}

\clearpage

\section{Supporting Figures}

\begin{figure}[htbp]
  \centering
  \includegraphics[width=0.6\textwidth]{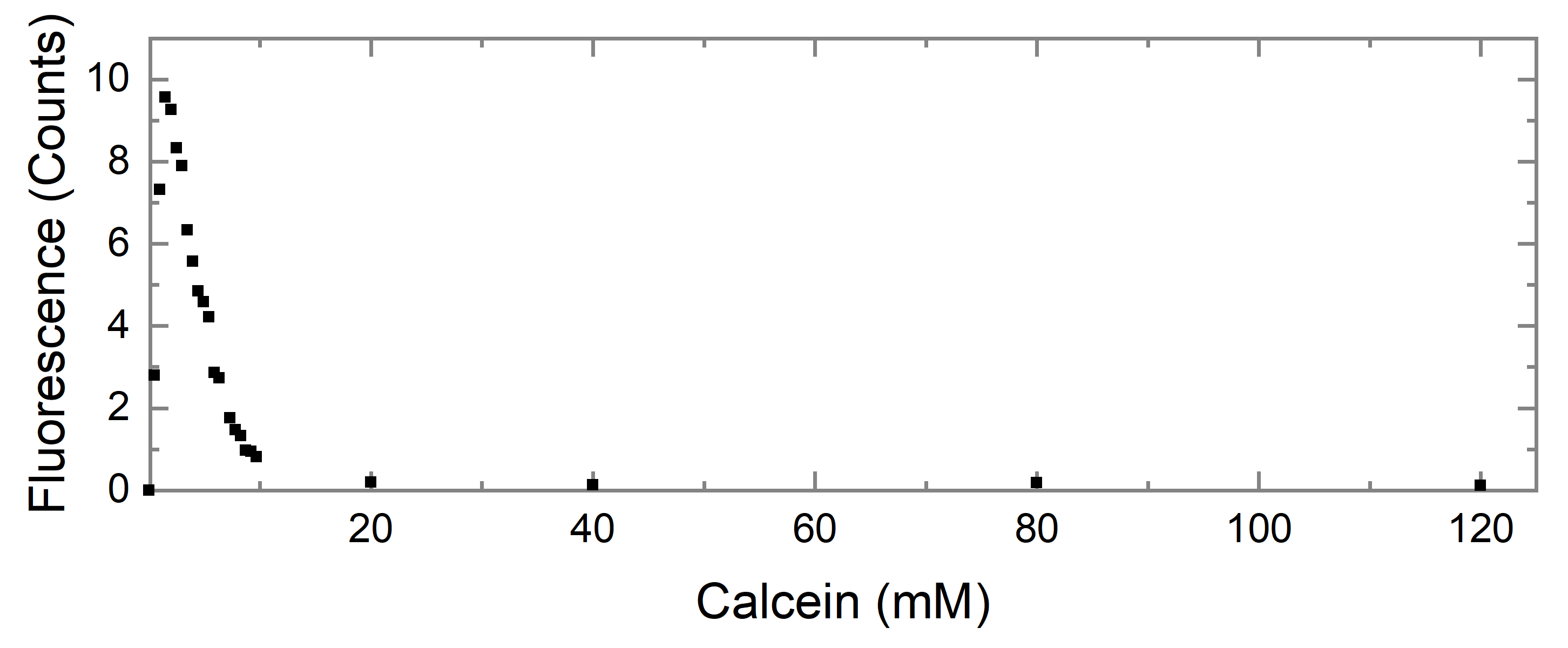}
  \caption{Calcein fluorescence in 0.3 M NaOH as function of Calcein concentraion. Above the concentration of 3 mM the fluorescence intensity of the calcein drops because of self-quenching of the dye.}
  \label{fig:CalceinConc}
\end{figure}

\begin{figure}[htbp]
  \centering
  \includegraphics[width=0.6\textwidth]{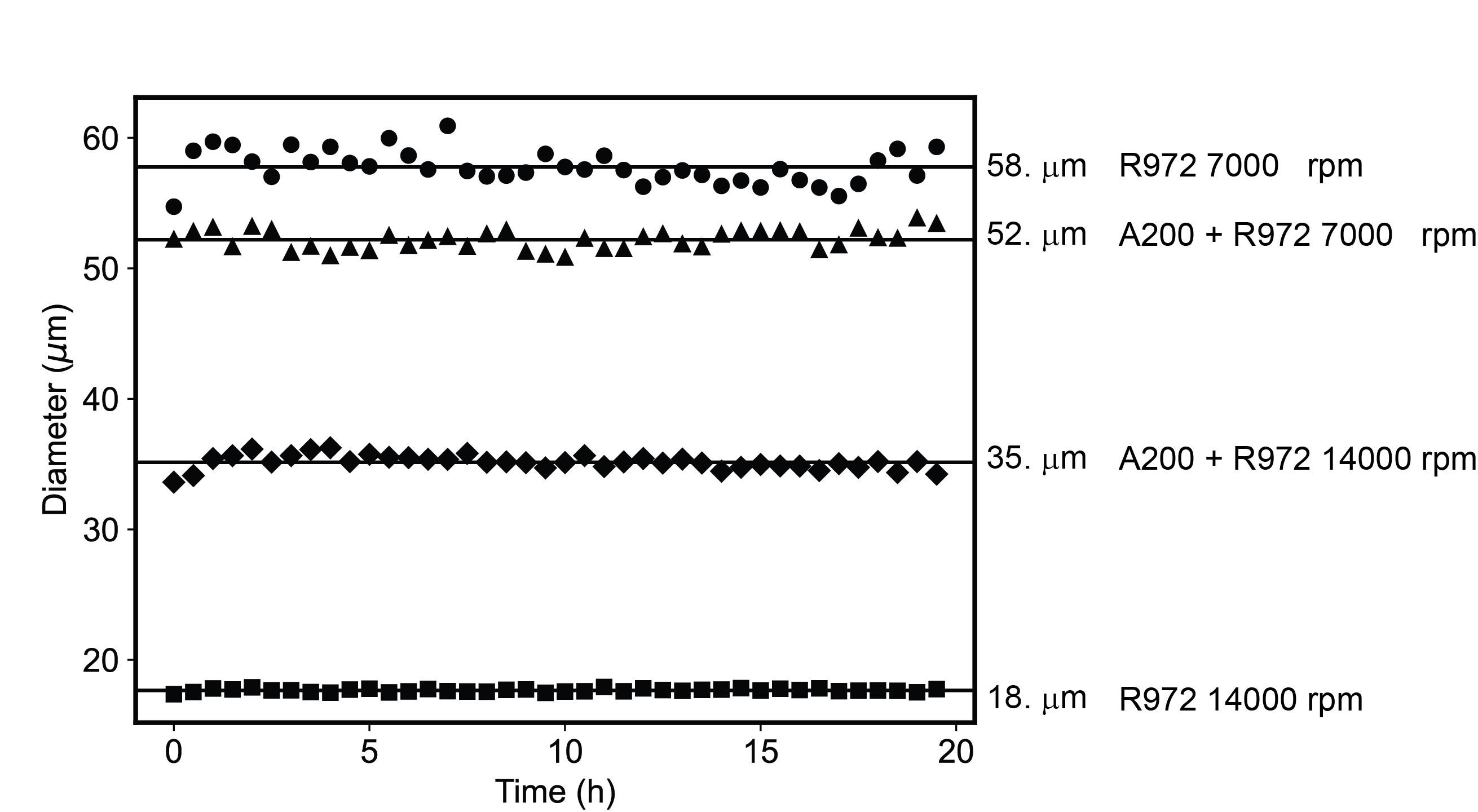}
  \caption{The stability of rehydrated antibubbles was evaluated by monitoring their size variation over a 20-hour time period. No noticeable alteration in diameter was observed for any of the investigated formulations. A minimum of n = 190 data points were assessed for each formulation.}
  \label{fig:Stability}
\end{figure}

\begin{figure}[htbp]
  \centering
  \begin{subfigure}[t]{0.45\textwidth}
    \centering
    \includegraphics[width=\textwidth]{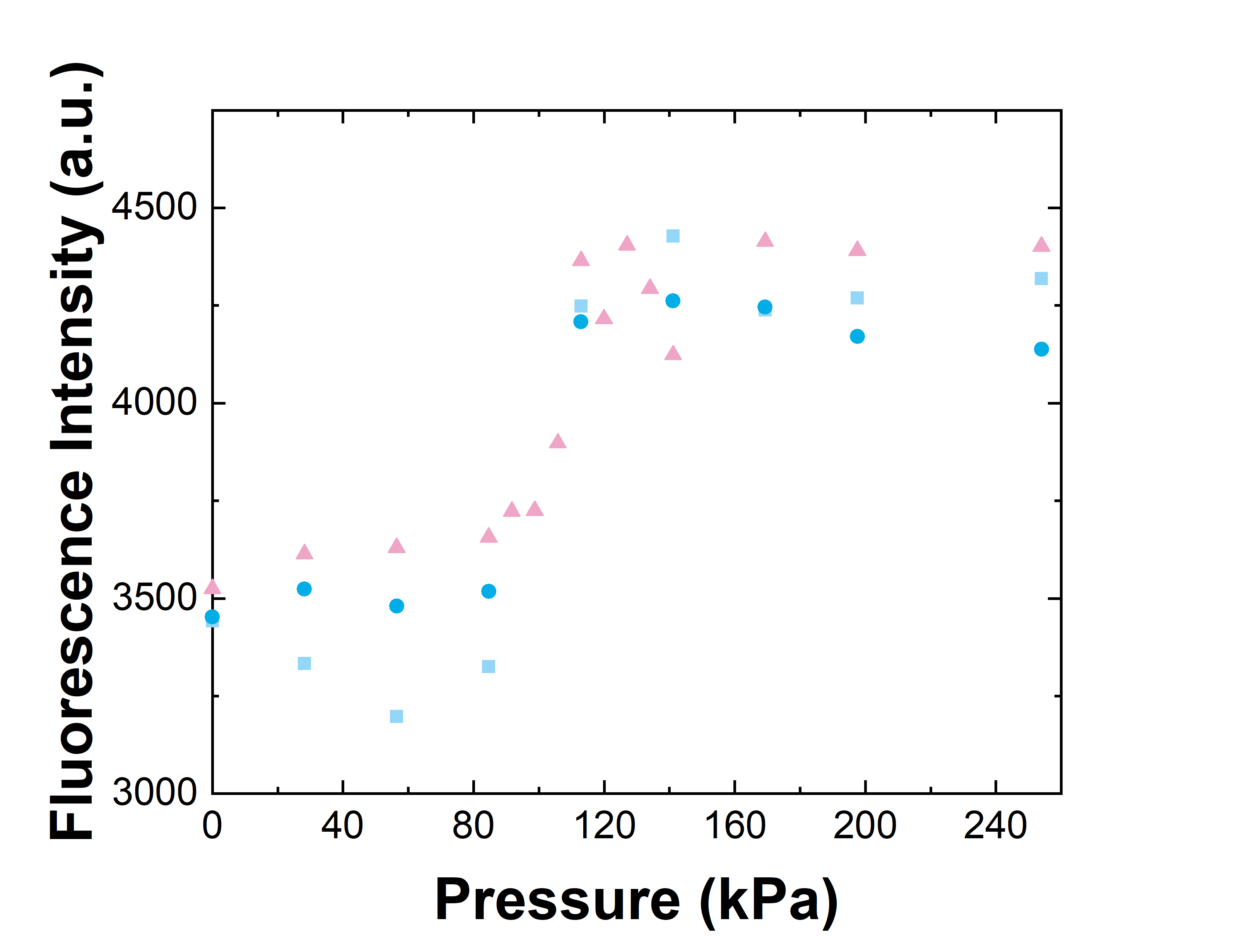}
    \caption{Aerosil 200 and Aerosil R972 - 7000 rpm}
  \end{subfigure}
  \hfill
  \begin{subfigure}[t]{0.45\textwidth}
    \centering
    \includegraphics[width=\textwidth]{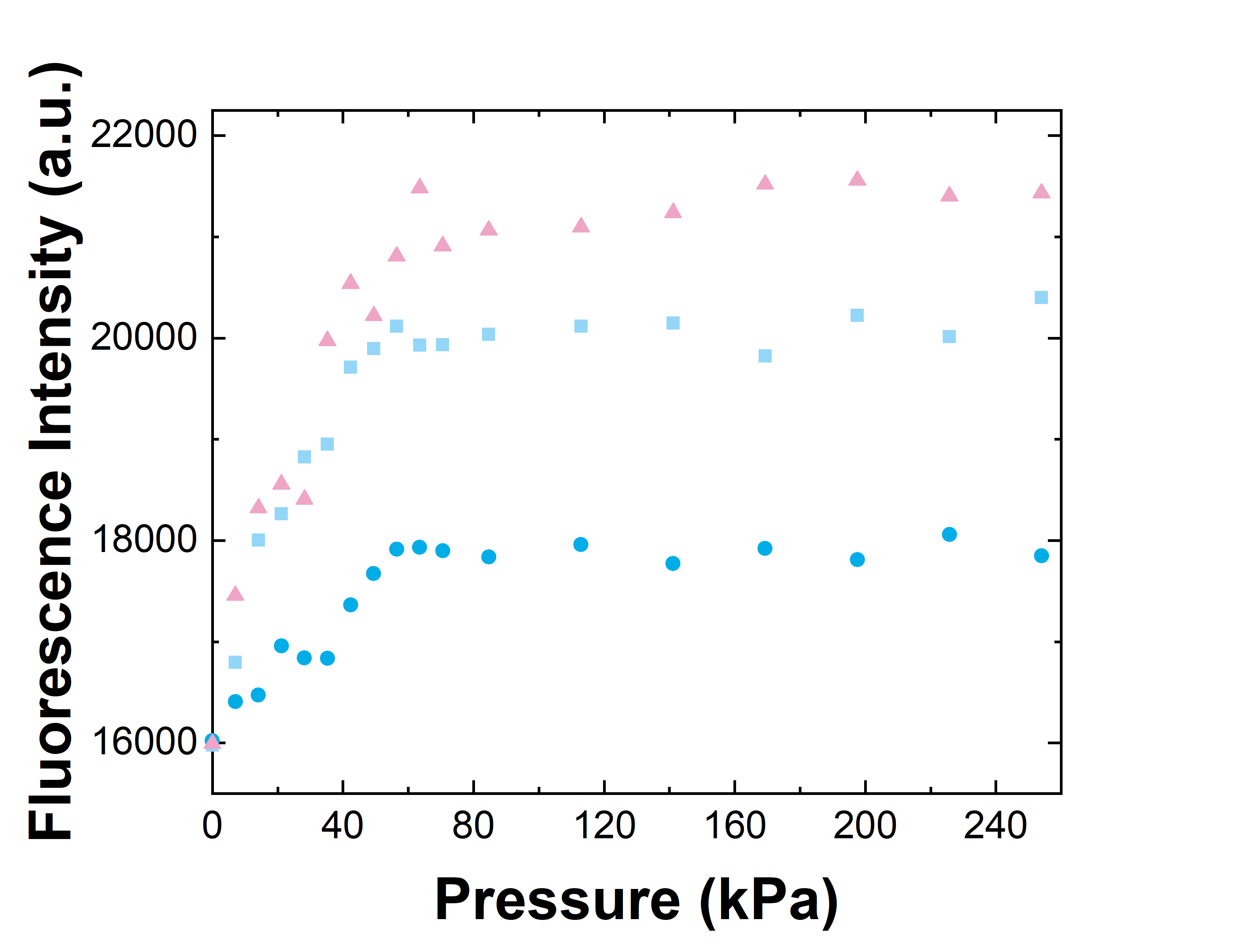}
    \caption{Aerosil 200 and Aerosil R972 - 14000 rpm}
  \end{subfigure}

  \vspace{0.5cm}

  \begin{subfigure}[t]{0.45\textwidth}
    \centering
    \includegraphics[width=\textwidth]{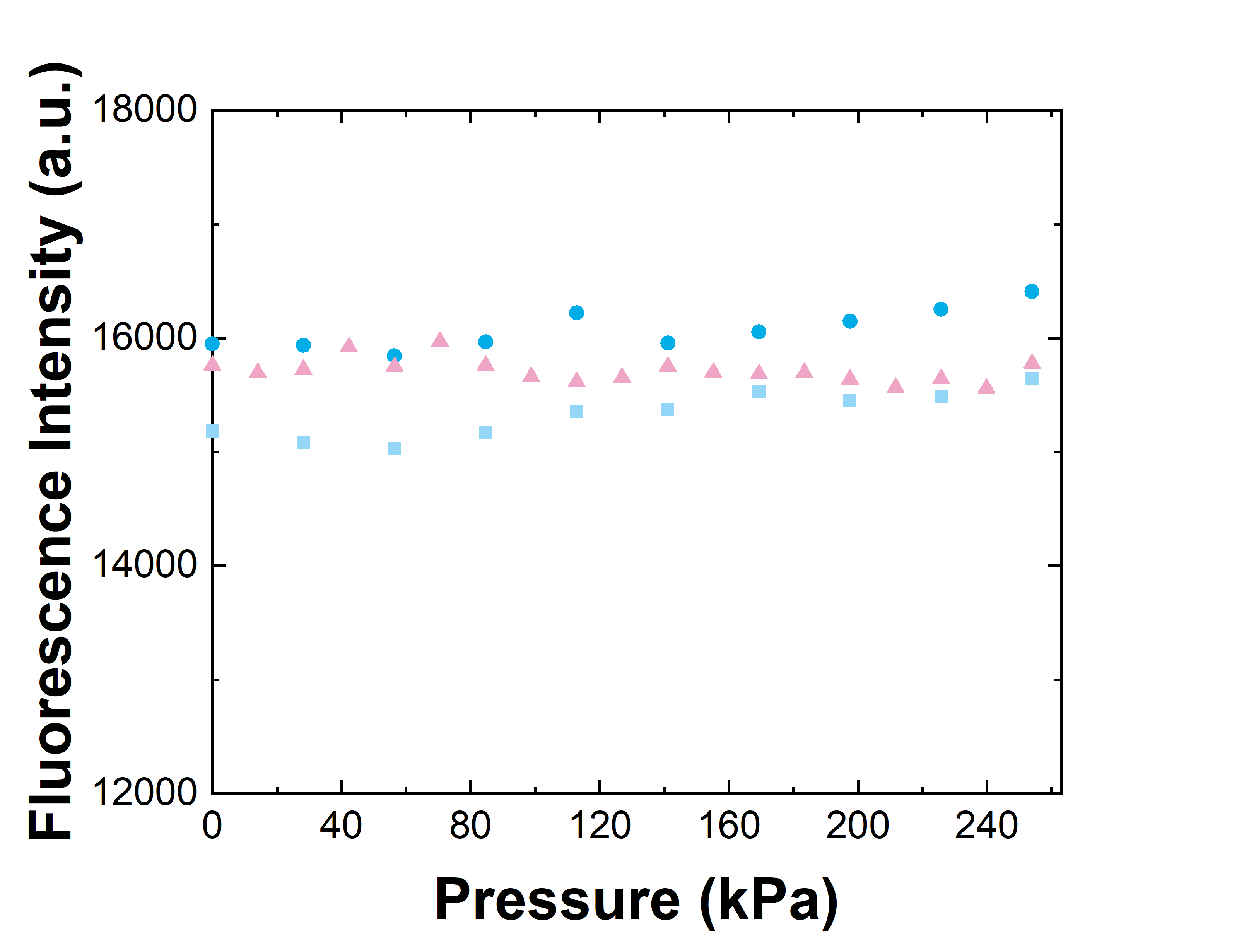}
    \caption{Aerosil R972 - 7000 rpm}
  \end{subfigure}
  \hfill
  \begin{subfigure}[t]{0.45\textwidth}
    \centering
    \includegraphics[width=\textwidth]{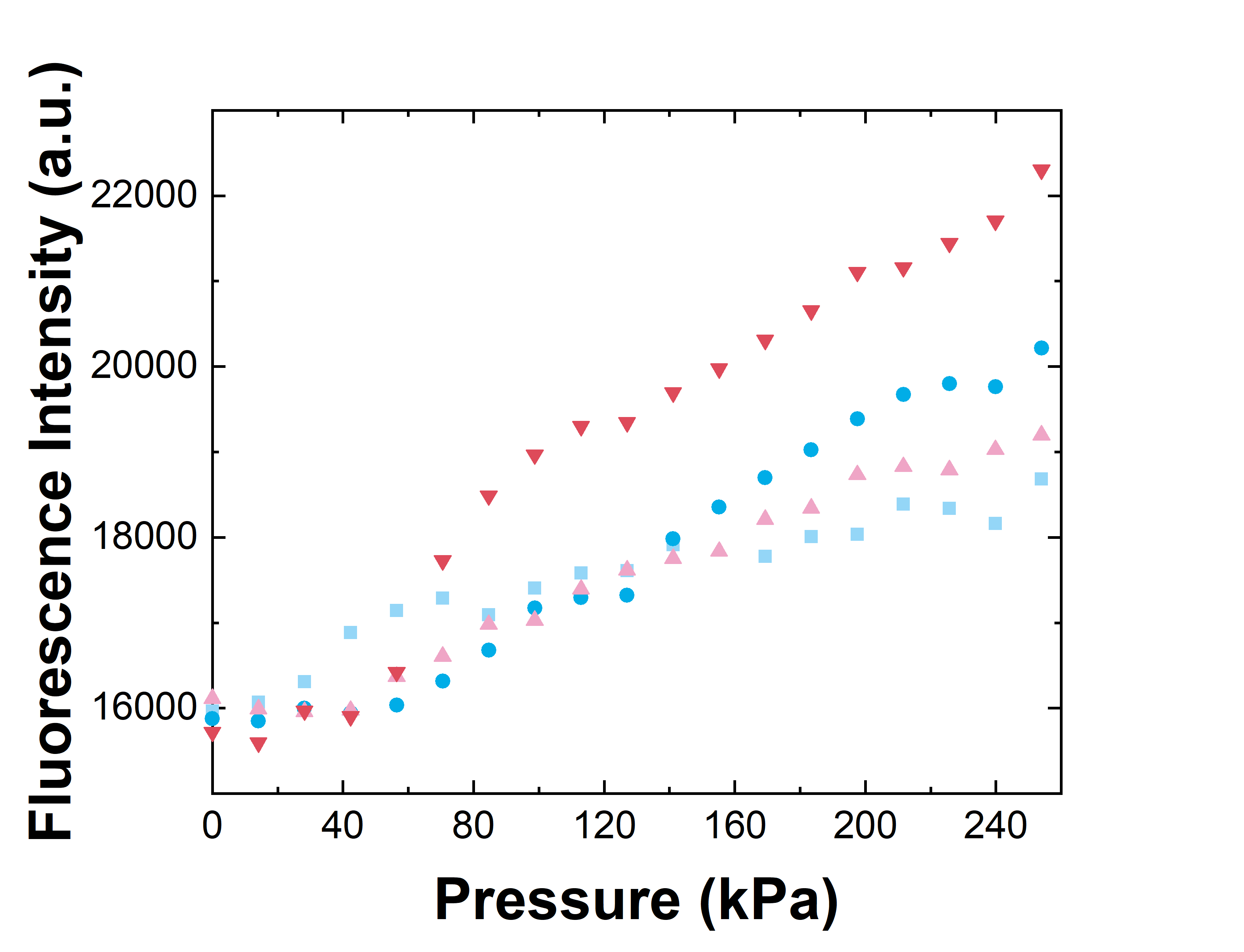}
    \caption{Aerosil R972 - 14000 rpm}
  \end{subfigure}
  \caption{Raw data for the bulk fluorescence measurements. Each panel shows the recorded fluorescence intensity (arbitrary units) as a function of pressure, reported as the peak hydrophone voltage (mVpp) measured in the cuvette. Panel (a) and (b) correspond to the hydrophilic formulations, panel (c) and (d) to the hydrophobic one (see main text). Peak-to-peak drive voltage amplitudes were converted to peak acoustic pressures in the main text using a calibration factor of \SI{1.411}{\kilo\pascal\per\milli\volt}, determined from hydrophone measurements. Different markers represent independently measured samples.}
  \label{fig:Si_Rawdata}
\end{figure}

\begin{figure}[htbp]
  \centering
  \includegraphics[width=0.6\textwidth]{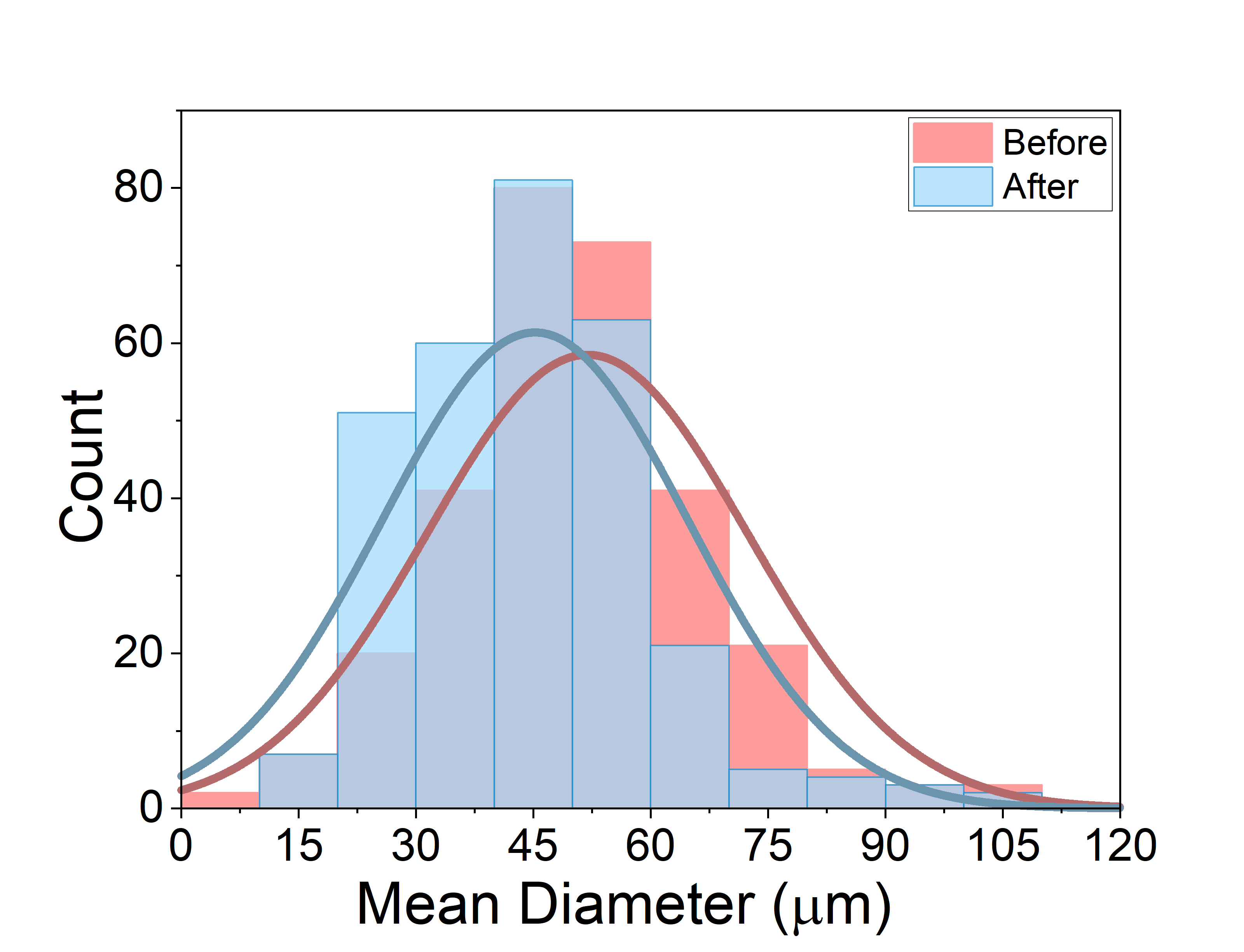}
  \caption{Size distribution for the Hydrophilic sample homogenized at 7000 rpm.}
  \label{fig:Size_dist_A200R97_7000rpm}
\end{figure}

\begin{figure}[htb]
    \centering
    \includegraphics[width=0.6\textwidth]{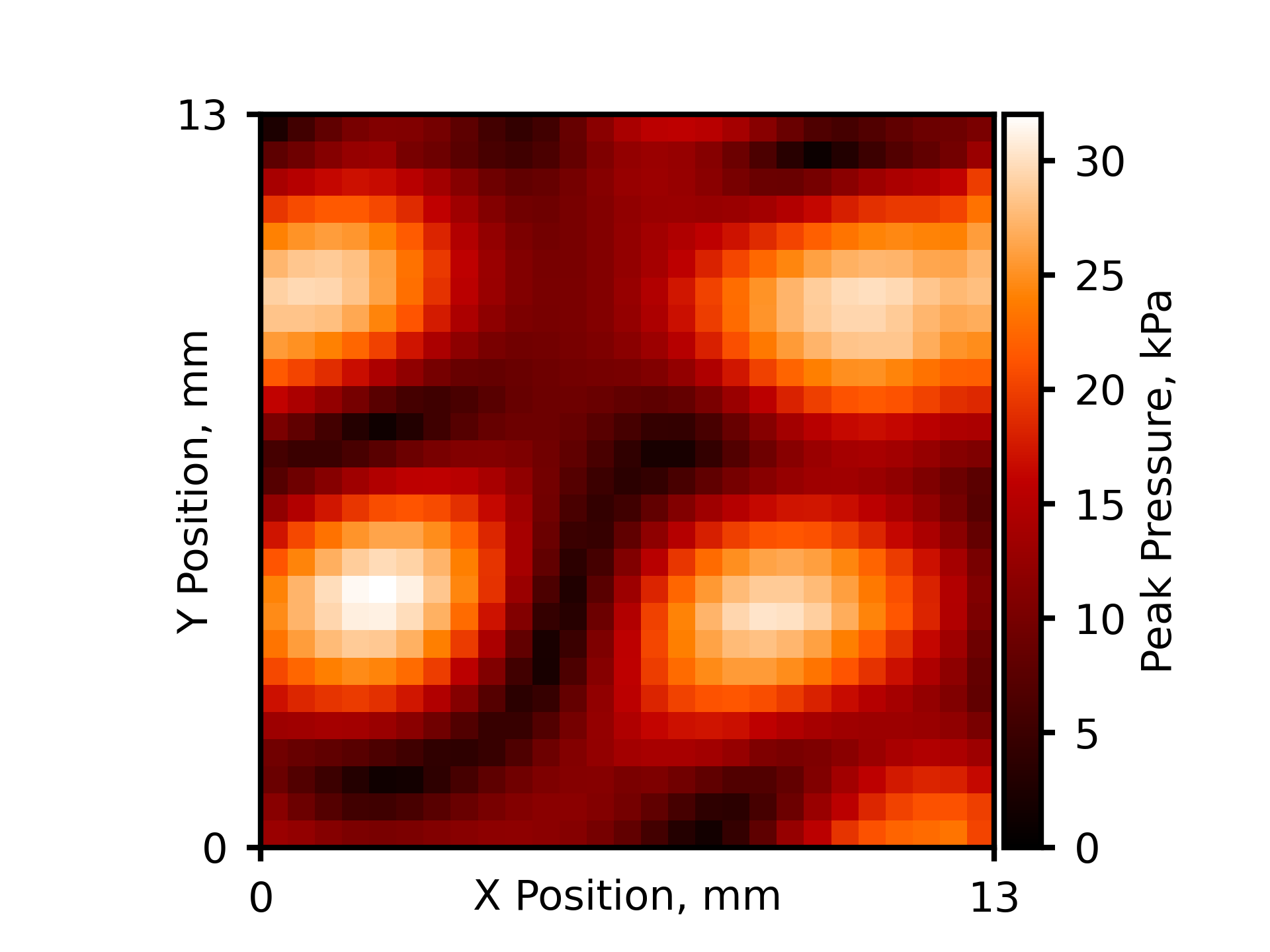}
    \caption{Measured pressure amplitude at 90.5 kHz within the sample chamber on the glass slide.}
    \label{fig:SI:glass_plate_pressure}
\end{figure}

\clearpage


\section{Supporting Videos}
\begin{description}
\item[Video S1] \hfill \\Confocal Z-stack of a $\sim$\SI{30}{\micro\meter}-diameter hydrophilic antibubble. Compared to Video S2, the encapsulated calcein concentration was lower so that the core is visible. Background fluorescence is not visible here because a lower exposure time and gain were needed to observe the inner core.
\item[Video S2] \hfill \\ 
Confocal Z-stack of a $\sim$\SI{30}{\micro\meter}-diameter hydrophobic antibubble. Because of the high concentration of calcein loaded in the antibubble, the core is self-quenched and no fluorescence is visible internally. Free calcein in the solution and bound to free-floating silica is visible because of the long exposure and gain settings on the microscope.
\item[Video S3] \hfill \\ Dynamics of 3 antibubbles during ultrasonic radiation at 90.5 kHz, recorded at 120000 fps.
\item[Video S4] \hfill \\ Dynamics of 2 antibubbles during ultrasonic radiation at 90.5 kHz, recorded at 120000 fps.
\end{description}

\clearpage

\medskip

\bibliographystyle{MSP}
\bibliography{References}

\begin{thebibliography}{10}
\providecommand{\url}[1]{\texttt{#1}}
\providecommand{\urlprefix}{URL }

\bibitem{melde2018acoustic}
K.~Melde, E.~Choi, Z.~Wu, S.~Palagi, T.~Qiu, P.~Fischer,
\newblock \emph{Advanced Materials} \textbf{2018}, \emph{30}, 3 1704507.

\bibitem{aubert2019spatial}
S.~Aubert, M.~Bezagu, A.~C. Spivey, S.~Arseniyadis,
\newblock \emph{Nature Reviews Chemistry} \textbf{2019}, \emph{3}, 12 706.

\bibitem{zhou2022advances}
Y.~Zhou, G.~Liu, S.~Guo,
\newblock \emph{Journal of Materials Chemistry B} \textbf{2022}, \emph{10}, 21
  3947.

\bibitem{blackmore2023ultrasound}
D.~G. Blackmore, D.~Razansky, J.~G{\"o}tz,
\newblock \emph{Neuron} \textbf{2023}, \emph{111}, 8 1174.

\bibitem{athanassiadis2021ultrasound}
A.~G. Athanassiadis, Z.~Ma, N.~Moreno-Gomez, K.~Melde, E.~Choi, R.~Goyal,
  P.~Fischer,
\newblock \emph{Chemical Reviews} \textbf{2021}, \emph{122}, 5 5165.

\bibitem{li2022recent}
J.~Li, Y.~Ma, T.~Zhang, K.~K. Shung, B.~Zhu,
\newblock \emph{BME Frontiers} \textbf{2022}.

\bibitem{melde2016holograms}
K.~Melde, A.~G. Mark, T.~Qiu, P.~Fischer,
\newblock \emph{Nature} \textbf{2016}, \emph{537}, 7621 518.

\bibitem{lin2021magnetism}
F.-C. Lin, Y.~Xie, T.~Deng, J.~I. Zink,
\newblock \emph{Journal of the American Chemical Society} \textbf{2021},
  \emph{143}, 16 6025.

\bibitem{wang2017external}
Y.~Wang, D.~S. Kohane,
\newblock \emph{Nature Reviews Materials} \textbf{2017}, \emph{2}, 6 1.

\bibitem{dholakia:2020}
K.~Dholakia, B.~W. Drinkwater, M.~{Ritsch-Marte},
\newblock \emph{Nature Reviews Physics} \textbf{2020}, \emph{2}, 9 480.

\bibitem{rwei2017ultrasound}
A.~Y. Rwei, J.~L. Paris, B.~Wang, W.~Wang, C.~D. Axon, M.~Vallet-Reg{\'\i},
  R.~Langer, D.~S. Kohane,
\newblock \emph{Nature biomedical engineering} \textbf{2017}, \emph{1}, 8 644.

\bibitem{delaney2022making}
L.~J. Delaney, S.~Isguven, J.~R. Eisenbrey, N.~J. Hickok, F.~Forsberg,
\newblock \emph{Materials Advances} \textbf{2022}.

\bibitem{xu2021employing}
N.~Xu, Z.~Song, M.-Z. Guo, L.~Jiang, H.~Chu, C.~Pei, P.~Yu, Q.~Liu, Z.~Li,
\newblock \emph{Cement and Concrete Composites} \textbf{2021}, \emph{118}
  103951.

\bibitem{song2022influence}
Z.~Song, N.~Xu, L.~Yu, M.-Z. Guo,
\newblock \emph{Journal of Building Engineering} \textbf{2022}, \emph{62}
  105413.

\bibitem{nele2020ultrasound}
V.~Nele, C.~E. Schutt, J.~P. Wojciechowski, W.~Kit-Anan, J.~J. Doutch, J.~P.
  Armstrong, M.~M. Stevens,
\newblock \emph{Advanced Materials} \textbf{2020}, \emph{32}, 7 1905914.

\bibitem{stride2019nucleation}
E.~Stride, C.~Coussios,
\newblock \emph{Nature Reviews Physics} \textbf{2019}, \emph{1}, 8 495.

\bibitem{afadzi2012effect}
M.~Afadzi, C.~d.~L. Davies, Y.~H. Hansen, T.~Johansen, {\O}.~K. Standal,
  R.~Hansen, S.-E. M{\aa}s{\o}y, E.~A. Nilssen, B.~Angelsen,
\newblock \emph{Ultrasound in medicine \& biology} \textbf{2012}, \emph{38}, 3
  476.

\bibitem{el2021molecular}
F.~El~Hajj, P.~F. Fuchs, W.~Urbach, M.~Nassereddine, S.~Hamieh, N.~Taulier,
\newblock \emph{Langmuir} \textbf{2021}, \emph{37}, 13 3868.

\bibitem{couture2011ultrasound}
O.~Couture, M.~Faivre, N.~Pannacci, A.~Babataheri, V.~Servois, P.~Tabeling,
  M.~Tanter,
\newblock \emph{Medical physics} \textbf{2011}, \emph{38}, 2 1116.

\bibitem{chomas:2000}
J.~E. Chomas, P.~A. Dayton, D.~May, J.~Allen, A.~Klibanov, K.~Ferrara,
\newblock \emph{Applied Physics Letters} \textbf{2000}, \emph{77}, 7 1056.

\bibitem{chomas:2001}
J.~E. Chomas, P.~A. Dayton, D.~J. May, K.~W. Ferrara,
\newblock \emph{Journal of Biomedical Optics} \textbf{2001}, \emph{6}, 2 141.

\bibitem{shapiro:2014}
M.~G. Shapiro, P.~W. Goodwill, A.~Neogy, M.~Yin, F.~S. Foster, D.~V. Schaffer,
  S.~M. Conolly,
\newblock \emph{Nature Nanotechnology} \textbf{2014}, \emph{9}, 4 311.

\bibitem{chandan:2020}
R.~Chandan, S.~Mehta, R.~Banerjee,
\newblock \emph{ACS Biomaterials Science \& Engineering} \textbf{2020},
  \emph{6}, 9 4731.

\bibitem{usfoodanddrugadministration:2023}
{US Food and Drug Administration},
\newblock Marketing clearance of diagnostic ultrasound systems and transducers:
  {{Guidance}} for industry and {{Food}} and {{Drug Administration}} staff,
\newblock Technical Report FDA-2017-D-5372, \textbf{2023}.

\bibitem{thanhnguyen:2017}
T.~Thanh~Nguyen, Y.~Asakura, S.~Koda, K.~Yasuda,
\newblock \emph{Ultrasonics Sonochemistry} \textbf{2017}, \emph{39} 301.

\bibitem{duck:2007}
F.~A. Duck,
\newblock \emph{Progress in Biophysics and Molecular Biology} \textbf{2007},
  \emph{93}, 1 176.

\bibitem{baresch2020acoustic}
D.~Baresch, V.~Garbin,
\newblock \emph{Proceedings of the National Academy of Sciences} \textbf{2020},
  \emph{117}, 27 15490.

\bibitem{kar2022wearable}
A.~Kar, N.~Ahamad, M.~Dewani, L.~Awasthi, R.~Patil, R.~Banerjee,
\newblock \emph{Biomaterials} \textbf{2022}, \emph{283} 121435.

\bibitem{poortinga2011long}
A.~T. Poortinga,
\newblock \emph{Langmuir} \textbf{2011}, \emph{27}, 6 2138.

\bibitem{silpe2013generation}
J.~E. Silpe, J.~K. Nunes, A.~T. Poortinga, H.~A. Stone,
\newblock \emph{Langmuir} \textbf{2013}, \emph{29}, 28 8782.

\bibitem{jiang2023high}
J.~Jiang, A.~T. Poortinga, Y.~Liao, T.~Kamperman, C.~H. Venner, C.~W. Visser,
\newblock \emph{Advanced Materials} \textbf{2023}, 2208894.

\bibitem{poortinga2013micron}
A.~T. Poortinga,
\newblock \emph{Colloids and Surfaces A: Physicochemical and Engineering
  Aspects} \textbf{2013}, \emph{419} 15.

\bibitem{kotopoulis2022formulation}
S.~Kotopoulis, C.~Lam, R.~Haugse, S.~Snipstad, E.~Murvold, T.~Jouleh, S.~Berg,
  R.~Hansen, M.~Popa, E.~Mc~Cormack, et~al.,
\newblock \emph{Ultrasonics Sonochemistry} \textbf{2022}, 105986.

\bibitem{vitry2019controlling}
Y.~Vitry, S.~Dorbolo, J.~Vermant, B.~Scheid,
\newblock \emph{Advances in colloid and interface science} \textbf{2019},
  \emph{270} 73.

\bibitem{zia2022advances}
R.~Zia, A.~Nazir, A.~T. Poortinga, C.~F. van Nostrum,
\newblock \emph{Advances in Colloid and Interface Science} \textbf{2022},
  102688.

\bibitem{lentacker2009drug}
I.~Lentacker, S.~C. De~Smedt, N.~N. Sanders,
\newblock \emph{Soft Matter} \textbf{2009}, \emph{5}, 11 2161.

\bibitem{borden2018reverse}
M.~A. Borden, K.-H. Song,
\newblock \emph{Advances in colloid and interface science} \textbf{2018},
  \emph{262} 39.

\bibitem{EvonikAerosil}
Evonik,
\newblock \emph{AEROSIL® fumed silica - Product overview},
\newblock Evonik, Evonik Resource Efficiency GmbH Business Line Silica
  Rodenbacher Chaussee 4 63457 Hanau Germany, \textbf{2018}.

\bibitem{kotopoulis:2015}
S.~Kotopoulis, K.~Johansen, O.~Gilja, A.~Poortinga, M.~Postema,
\newblock \emph{Acta Physica Polonica A} \textbf{2015}, \emph{127}, 1 99.

\bibitem{noble2014digital}
M.~L. Noble, P.~D. Mourad, B.~D. Ratner,
\newblock \emph{Biomaterials science} \textbf{2014}, \emph{2}, 6 893.

\bibitem{xia2016ultrasound}
H.~Xia, Y.~Zhao, R.~Tong,
\newblock \emph{Therapeutic Ultrasound} \textbf{2016}, 365--384.

\bibitem{kennedy2016sequential}
S.~Kennedy, J.~Hu, C.~Kearney, H.~Skaat, L.~Gu, M.~Gentili, H.~Vandenburgh,
  D.~Mooney,
\newblock \emph{Biomaterials} \textbf{2016}, \emph{75} 91.

\bibitem{moncion2018sequential}
A.~Moncion, M.~Lin, O.~D. Kripfgans, R.~T. Franceschi, A.~J. Putnam, M.~L.
  Fabiilli,
\newblock \emph{Ultrasound in medicine \& biology} \textbf{2018}, \emph{44}, 11
  2323.

\bibitem{aliabouzar2021stable}
M.~Aliabouzar, O.~D. Kripfgans, W.~Y. Wang, B.~M. Baker, J.~B. Fowlkes, M.~L.
  Fabiilli,
\newblock \emph{Ultrasonics sonochemistry} \textbf{2021}, \emph{72} 105430.

\bibitem{sijl:2011}
J.~Sijl, M.~Overvelde, B.~Dollet, V.~Garbin, N.~{de Jong}, D.~Lohse,
  M.~Versluis,
\newblock \emph{The Journal of the Acoustical Society of America}
  \textbf{2011}, \emph{129}, 4 1729.

\bibitem{fesce:1994}
R.~Fesce, F.~Grohovaz, F.~Valtorta, J.~Meldolesi,
\newblock \emph{Trends in Cell Biology} \textbf{1994}, \emph{4}, 1 1.

\bibitem{freiberg:2004}
S.~Freiberg, X.~X. Zhu,
\newblock \emph{International Journal of Pharmaceutics} \textbf{2004},
  \emph{282}, 1 1.

\bibitem{timko:2014}
B.~P. Timko, M.~Arruebo, S.~A. Shankarappa, J.~B. McAlvin, O.~S. Okonkwo,
  B.~Mizrahi, C.~F. Stefanescu, L.~Gomez, J.~Zhu, A.~Zhu, J.~Santamaria,
  R.~Langer, D.~S. Kohane,
\newblock \emph{Proceedings of the National Academy of Sciences} \textbf{2014},
  \emph{111}, 4 1349.

\bibitem{mardanighahfarokhi:2020}
V.~Mardani~Ghahfarokhi, P.~P. Pescarmona, G.-J.~W. Euverink, A.~T. Poortinga,
\newblock \emph{Colloids and Interfaces} \textbf{2020}, \emph{4}, 3 40.

\bibitem{schindelin2012fiji}
J.~Schindelin, I.~Arganda-Carreras, E.~Frise, V.~Kaynig, M.~Longair,
  T.~Pietzsch, S.~Preibisch, C.~Rueden, S.~Saalfeld, B.~Schmid, et~al.,
\newblock \emph{Nature methods} \textbf{2012}, \emph{9}, 7 676.

\end{thebibliography}

\end{document}